\documentclass[a4paper,12pt]{article}

\usepackage{graphics}
\usepackage{latexsym}
\usepackage{amsmath}
\usepackage{amssymb}
\newcommand{\ap}{\alpha^{\prime}}
\newcommand{\cA}{\mathcal{A}}

\newcommand{\cF}{\mathcal{F}}

\newcommand{\cH}{\mathcal{H}}

\newcommand{\cL}{\mathcal{L}}

\newcommand{\cN}{\mathcal{N}}
\newcommand{\cO}{\mathcal{O}}

\newcommand{\cQ}{\mathcal{Q}}
\newcommand{\cR}{\mathcal{R}}

\newcommand{\cT}{\mathcal{T}}

\newcommand{\cV}{\mathcal{V}}

\newcommand{\cZ}{\mathcal{Z}}
\newcommand{\zetto}{\mathbb{Z}}
\newcommand{\aaru}{\mathbb{R}}
\newcommand{\shii}{\mathbb{C}}
\newcommand{\gh}{\#_{\mathrm{gh}}}
\newcommand{\pic}{\#_{\mathrm{pic}}}

\newcommand{\llk}{\langle\!\langle}
\newcommand{\rrk}{\rangle\!\rangle}
\newcommand{\bllk}{\biggl\langle\!\!\!\biggl\langle}
\newcommand{\brrk}{\biggr\rangle\!\!\!\biggr\rangle}

\newcommand{\Qev}{\cQ_{\mathrm{even}}}
\newcommand{\Qod}{\cQ_{\mathrm{odd}}}

\newcommand{\sectiono}[1]{\section{#1}\setcounter{equation}{0}}

\begin{document}
\begin{titlepage}
\thispagestyle{empty}
\begin{flushright}
UT-03-15 \\
hep-th/0305103 \\
May, 2003 
\end{flushright}

\vskip 1.5 cm

\begin{center}
\noindent{\textbf{\LARGE{ Level-Expansion Analysis in 
\vspace{0.5cm}\\ NS Superstring Field Theory Revisited }}} 
\vskip 1.5cm
\noindent{\large{Kazuki Ohmori}\footnote{E-mail: ohmori@hep-th.phys.s.u-tokyo.ac.jp}}\\ 
\vspace{1cm}
\noindent{\small{\textit{Department of Physics, Faculty of Science, University of 
Tokyo}} \\ \vspace{2mm}
\small{\textit{Hongo 7-3-1, Bunkyo-ku, Tokyo 113-0033, Japan}}}
\end{center}
\vspace{1cm}
\begin{abstract}
We study the level-expansion structure of the NS string field theory actions, 
mainly focusing on the modified (\textit{i.e.} 0-picture in the NS sector) 
cubic superstring field theory. This theory has a non-trivial structure 
already at the quadratic level due to presence of the picture-changing operator. 
It is explicitly shown how the usual Maxwell and tachyon actions can be obtained 
after integrating out the auxiliary fields. 
We then discuss the reality of the action in the CFT language for all of modified cubic, 
Witten's cubic and Berkovits' non-polynomial theories. 
The tachyon condensation problems in modified cubic theory are re-examined. 
We also carry out level truncation analysis in 
vacuum superstring field theory proposed in our previous paper, 
and find some difficulties in both of cubic and non-polynomial formulations. 
\end{abstract}
\end{titlepage}
\newpage
\baselineskip 6mm


\sectiono{Introduction}\label{sec:introduction}
The construction of covariant open superstring field theory 
based on the RNS formalism~\cite{FMS}
has been a long-standing problem due to the complications coming from the concept of `picture'. 
In Witten's original proposal~\cite{Witten2} for cubic superstring field theory, 
which was the natural extension of his bosonic cubic string field theory~\cite{Witten1}, 
the NS string field was taken to be in the `natural' $-1$-picture. However, it turned out that 
this theory suffered from contact-term divergences at the tree-level caused by the 
colliding picture-changing operators~\cite{Wendt}. 
About 10 years later, Berkovits found a way to construct a gauge-invariant NS open string field 
theory action without making use of the picture-changing operators~\cite{sP}. 
However, it has been realized that  
it is impossible to include the Ramond (R) sector string field in the action 
in a manifestly ten-dimensional Lorentz covariant 
manner without introducing the picture-changing operations~\cite{Ramond}. 
Furthermore, its ten-dimensional supersymmetric structure still 
remains unclear.\footnote{By using the hybrid formalism the \textit{four}-dimensional $\cN =1$ 
super-Poincar\'{e} invariance can be made manifest. And it was recently 
discussed how to deal with the GSO($-$) sector in the hybrid formalism~\cite{BT}.}

Before Berkovits' discovery, in 1990 more conservative method of modifying 
Witten's cubic theory had been proposed~\cite{AMZ1,AMZ2,PTY}. 
There, the NS string field was defined to carry picture number 0 so that 
the quadratic vertex had the same picture-changing operator insertion as 
the cubic vertex. 
In spite of containing the picture-changing operators in the action, 
it was shown~\cite{AMZ2,PTY} that this theory is 
free from contact-term divergence problems. 
With the help of picture-changing operators, 
we are able not only to include the R sector string field 
of picture number $-\frac{1}{2}$ in a ten-dimensional Lorentz covariant manner, 
but also to construct the $d=10$, $\cN =1$ spacetime supersymmetry generator.  
However, a subtle problem 
regarding the picture-changing operator still remains: 
The linearized equation of motion $Y_{-2}Q_BA=0$ for the NS string field 
can differ from the usual one $Q_BA=0$ because $Y_{-2}$ has a 
non-trivial kernel. But 
since the picture-changing operator is inserted at the open string midpoint ($\pm i$ in the UHP 
representation), $Y_{-2}Q_BA=0$ gives the same result as $Q_BA=0$ as long as $A$ is restricted 
to being in the \textit{finite}-dimensional Fock space. 
Hence it is conceivable that the level truncation procedure provides 
an \textit{ad hoc} way of regularizing\footnote{In the context of (discrete) 
Moyal formulation of string field theory (MSFT), a more rigorous way 
to regularize (or cut-off) Witten's cubic string field theory has been proposed~\cite{MSFT}.}
the problem of non-trivial kernel of $Y_{-2}$,
although the explicit truncation of the states in the kernel of $Y_{-2}$ 
in the \textit{full} theory would 
ruin the associativity of the $*$-product~(see \textit{e.g.} \cite{Ramond,0105230}). 
One of the aims of this paper is to see to what extent the level-truncated 
modified cubic superstring field theory can reproduce the 
structure expected of open superstring theory: Action for the low-lying fields, 
open string tachyon condensation\footnote{For earlier studies on tachyon 
condensation, see~\cite{BH}}, etc. 
\smallskip

This work is also motivated by the desire to investigate the proposed form of 
vacuum superstring field theory~\cite{0208009} within the level truncation 
scheme, because no exact D-brane solutions have been found in this theory so far. 
\smallskip

Besides, we have found that there have been very few reports on how the reality 
of the action is guaranteed by imposing appropriate conditions on the string fields. 
We will answer this question using the CFT method for all three proposals for 
superstring field theory. 
\medskip

This paper is organized as follows. In section~\ref{sec:massless} 
we calculate the cubic superstring field theory action for the NS($+$) massless sector, 
and show that, even though the $c_0L_0^{\mathrm{m}}$ term is absent, 
the correct Maxwell lagrangian can be reproduced after integrating out the 
auxiliary fields. 
In section~\ref{sec:GSO-} we discuss how to include the NS($-$) states 
in the action, paying attention to the problem of fixing sign ambiguities. 
Section~\ref{sec:reality} is devoted to the discussion about the reality conditions 
on the string fields. 
In section~\ref{sec:tachyonpotential} we re-examine the non-trivial vacuum solutions 
previously obtained 
in the non--GSO-projected~\cite{ABKM1} and GSO-projected~\cite{AMZssb} theories 
in the level truncation scheme, and present a (new) space-dependent kink solution 
of codimension 1 on a non-BPS D-brane at the lowest level. 
In section~\ref{sec:VSFTsolutions} we perform the level-truncation analysis 
in vacuum superstring field theory. 
We summarize our results in section~\ref{sec:summary}. 
The explicit expressions for the component action and technical 
remarks about the computations involving $X^{\mu}$ are collected in Appendices.

\sectiono{NS($+$) Massless Sector}\label{sec:massless}
We begin with the detailed study of how the massless gauge field, which belongs to 
the NS($+$) sector, is described in the modified cubic superstring field theory. 
Some results have already been shown in the literature~\cite{UZ,ABGKM}, but since we are using 
different conventions from theirs, we will explicitly write them down. 
The action for the NS($+$) string field $A_+$ is given by~\cite{AMZ1,AMZ2,PTY}
\begin{equation}
S=\frac{1}{g_o^2}\left(\frac{1}{2\ap}\llk Y_{-2}|A_+,Q_BA_+\rrk +\frac{1}{3}\llk Y_{-2}|A_+,A_+*A_+\rrk
\right), \label{eq:ZA}
\end{equation}
where $g_o$ is the open string coupling constant and the 2- and 3-string vertices are defined 
as the following correlation functions on the upper half plane,
\begin{eqnarray}
& &\llk Y_{-2}|A_1,A_2\rrk =\lim_{z\to 0}\left\langle Y(i)Y(-i)I\circ A_1(z)\ A_2(z)
\right\rangle_{\mathrm{UHP}}, \label{eq:ZB} \\
& &\llk Y_{-2}|A_1,A_2*A_3\rrk =\left\langle Y(i)Y(-i)f^{(3)}_1\circ A_1(0)
f^{(3)}_2\circ A_2(0)f^{(3)}_3\circ A_3(0)\right\rangle_{\mathrm{UHP}}, \label{eq:ZC}
\end{eqnarray}
where $Y=c\partial\xi e^{-2\phi}$ is the inverse picture-changing operator\footnote{It is 
possible to employ the following `chiral' double-step inverse 
picture-changing operator~\cite{AMZ2,PTY} 
\begin{equation}
\cZ=-e^{-2\phi}-\frac{1}{5}c\partial\xi e^{-3\phi} G^{\mathrm{m}} \label{eq:AMZpicturechange}
\end{equation}
as $Y_{-2}$, instead of the `non-chiral' choice $Y(i)Y(-i)$. However, it is known that 
the cubic theory with $\cZ$ has some problematic features: For example, the massless part of 
the free action does not reproduce the conventional Maxwell action~\cite{UZ}, 
and the insertion of $\cZ$ breaks the twist symmetry of the string field theory 
action~\cite{Raeymaekers} while $Y(i)Y(-i)$ preserves it.  
We will therefore focus on the non-chiral choice in this paper.} 
and $f\circ A(z)$ denotes the conformal 
transform of the vertex operator $A(z)$ by the conformal map $f$. Concretely, for a 
primary field $A$ of conformal weight $h$ we have 
$f\circ A(z)=(f^{\prime}(z))^h A(f(z))$. 
The conformal maps appearing in eqs.(\ref{eq:ZB}), (\ref{eq:ZC}) are 
\begin{equation}
I(z)=-\frac{1}{z}=h^{-1}(-h(z)), \quad f^{(3)}_k(z)=h^{-1}\left( e^{2\pi i\frac{k-2}{3}}
h(z)^{2/3}\right),\ \ (k=1,2,3) \label{eq:ZD}
\end{equation}
with 
\[ h(z)=\frac{1+iz}{1-iz}, \quad h^{-1}(z)=-i\frac{z-1}{z+1}. \] 
The correlation function is normalized as 
\begin{equation}
\left\langle \frac{1}{2}\partial^2c \partial c c(x) e^{-2\phi(y)} e^{ikX(w)}\right\rangle_{\mathrm{UHP}}
=(2\pi)^{10}\delta^{10}(k), \label{eq:ZE}
\end{equation}
and $(2\pi)^{10}\delta^{10}(0)\equiv V_{10}$ is the volume of the $(9+1)$-dimensional spacetime. 
The BRST operator $Q_B=\oint\frac{dz}{2\pi i}j_B(z)$ is nilpotent and acts as a graded derivation 
in the $*$-algebra. The picture-changing operator $Y_{-2}$ is BRST-invariant in the sense 
that $[Q_B,Y_{-2}]=\oint\frac{dz}{2\pi i}j_B(z)Y(i)Y(-i)=0$. The 3-string vertex, as well as 
the $n$-string vertices induced from the repeated use of the $*$-multiplication, satisfies the 
cyclicity relation 
\[ \llk Y_{-2}|A_1,A_2*A_3\rrk=\llk Y_{-2}|A_2,A_3*A_1\rrk=\llk Y_{-2}|A_3,A_1*A_2\rrk. \]
It then follows that the action~(\ref{eq:ZA}) is invariant under the gauge transformation 
\begin{equation}
\delta A_+=\frac{1}{\ap}Q_B\Lambda+A_+*\Lambda-\Lambda*A_+, \label{eq:gauge}
\end{equation}
where $\Lambda$ is an infinitesimal gauge transformation parameter. 

It is important to decide whether the overall multiplicative 
factor in front of the action~(\ref{eq:ZA}) should be positive or negative, because 
it cannot be absorbed by the redefinition of the \textit{real} string field. 
We cannot answer this question at this point, and 
it should be determined by looking at the sign of the kinetic term of the 
physical component field, as will be done. 
\medskip

For the superconformal ghost sector, we will entirely be working in the 
`fermionized' language\footnote{Alternatively, we can write the whole theory in terms of the 
$\beta\gamma$ ghosts, because this cubic theory is formulated within the ``small" Hilbert space 
(namely, without introducing the zero mode of $\xi$). For example, the inverse picture-changing 
operator can be written as $Y=c\delta^{\prime}(\gamma)$, with $\delta^{\prime}(\gamma)$ satisfying 
the property $\gamma\delta^{\prime}(\gamma)=-\delta(\gamma)$.} without referring to the original 
bosonic $\beta\gamma$-ghost system. But we would like to make a remark on 
the fermionization formula\footnote{The author would like to thank T. Kawano and I. Kishimoto 
for useful discussion on this point. } here. $\beta$ and $\gamma$ are fermionized as 
\begin{equation}
\beta =e^{-\phi}(-1)^{-N_F}\partial\xi,\quad \gamma=\eta e^{\phi}(-1)^{N_F}, \label{eq:ZF}
\end{equation}
where 
\[ N_F= \oint\frac{dz}{2\pi i}\left(-:bc:-:\xi\eta:+\sum_{a=0}^4:\psi_a^+\psi_a^-:\right) \]
and $ \psi_0^{\pm}=\frac{1}{\sqrt{2}}(\pm\psi_0+\psi_1)$, \ 
$\psi^{\pm}_a=\frac{1}{\sqrt{2}}(\psi_{2a}\pm i\psi_{2a+1})\quad (a=1,2,3,4)$. 
Since $N_F$ is an operator that 
counts the number (mod 2) of the world-sheet fermions $\psi^{\mu}, b, c, \xi$ and $\eta$, 
$\mathit{(-1)^{N_F}}$ \textit{anticommutes with them.} Thus $(-1)^{\pm N_F}$ is considered as 
a cocycle factor attached to $e^{\pm\phi}$ such that $e^{\pm\phi}(-1)^{\pm N_F}$ 
anticommutes with the world-sheet fermions as a whole. 
The existence of this cocycle factor is 
important because, if it were absent, the statistics of $\gamma$ and that of $\eta e^{\phi}$ 
would not agree. From the OPE 
\begin{eqnarray*}
:e^{q_1\phi(z)}::e^{q_2\phi(w)}:&=&(z-w)^{-q_1q_2}:e^{q_1\phi(z)}e^{q_2\phi(w)}: \\
&=&(z-w)^{-q_1q_2}\left(:e^{(q_1+q_2)\phi(w)}:+\cO(z-w)\right) 
\end{eqnarray*}
one finds that $e^{q_1\phi}$ and $e^{q_2\phi}$ naturally anticommute with each other when both 
$q_1$ and $q_2$ are odd integers. After all, we have found that $e^{q\phi}(-1)^{qN_F}$ with 
odd $q$ anticommutes with all of the fermions and $e^{q^{\prime}\phi}(-1)^{q^{\prime}N_F}$ 
with odd $q^{\prime}$, whereas $e^{q\phi}(-1)^{qN_F}$ with even $q=2n$ commutes with everything because 
$2nN_F$ in the NS sector 
is always an even integer. Therefore, we can abbreviate $e^{q\phi}(-1)^{qN_F}$ 
to $e^{q\phi}$, with the understanding that $e^{q\phi}$ should be treated as a fermion/boson 
when $q$ is odd/even, respectively. 
In fact, it appears that almost all the calculations in the literature have been performed 
using this `abbreviation rule'. Also in the rest of this paper we will simply 
regard $e^{q\phi}$ with odd $q$ as fermionic, instead of explicitly writing the cocycle 
factor $(-1)^{qN_F}$. 
\medskip

We define the ghost number current $j_{\mathrm{gh}}$ and the picture number current 
$j_{\mathrm{pic}}$ as 
\begin{equation}
j_{\mathrm{gh}}=-:bc:-:\xi\eta:,\quad j_{\mathrm{pic}}=:\xi\eta:-\partial\phi. \label{eq:ZG}
\end{equation}
Given the OPEs 
\begin{equation}
c(z)b(w)\sim\frac{1}{z-w},\quad \eta(z)\xi(w)\sim\frac{1}{z-w},\quad \phi(z)\phi(w)\sim
-\log(z-w), \label{eq:ZH}
\end{equation}
the assignment of the ghost and picture numbers for the ghost fields is found to be 
\[ \begin{array}{|c||c|c|c|c|c|}
\hline 
 & b & c & \xi & \eta & e^{q\phi} \\ \hline
\gh & -1 & 1 & -1 & 1 & 0 \\ \hline
\pic & 0 & 0 & 1 & -1 & q \\ \hline
\end{array}\ . \]
\medskip

The string field $A_+$ is defined to be a Grassmann-odd element in the state space 
of the 2-dimensional conformal field theory, consisting of states of ghost number 1 and 
picture number 0. At the massless level, it is expanded as 
\begin{eqnarray}
|A_+^{(0)}\rangle&=&A_+^{(0)}(0)|0\rangle, \nonumber \\ 
A_+^{(0)}(z)&=&\int\frac{d^{10}k}{(2\pi)^{10}}\biggl[\frac{i}{\sqrt{2\ap}}A^1_{\mu}(k)
c\partial X^{\mu}+A^2_{\mu}(k)\eta e^{\phi}\psi^{\mu} \label{eq:ZI} \\
& &{}+\frac{\sqrt{2\ap}}{2i}F_{\mu\nu}(k)c\psi^{\mu}\psi^{\nu}+iv(k)\partial c
+iw(k)c\partial\phi\biggr]e^{ikX}(z), \nonumber
\end{eqnarray}
where $|0\rangle$ denotes the $SL(2,\aaru)$-invariant vacuum. 
The reality condition on the string field 
implies the following reality conditions 
for the component fields (see section~\ref{sec:reality} for details), 
\begin{eqnarray}
& &A^1_{\mu}(k)^*=A^1_{\mu}(-k), \quad A^2_{\mu}(k)^*=A^2_{\mu}(-k), \quad 
F_{\mu\nu}(k)^*=F_{\mu\nu}(-k), \label{eq:ZN} \\
& &v(k)^*=v(-k), \quad w(k)^*=w(-k), \nonumber 
\end{eqnarray}
where ${}^*$ denotes the complex conjugation. 
\smallskip

We will now show the detailed calculations of the quadratic action, \textit{i.e.} 
rewriting the action~(\ref{eq:ZA}) in terms of the component fields appearing in~(\ref{eq:ZI}). 
In the Abelian case no cubic interactions among the massless 
fields~(\ref{eq:ZI}) survive due to the twist symmetry. 
First, we have to compute the action on $A_+^{(0)}$ of the BRST operator 
\begin{equation}
Q_B=\oint\frac{dz}{2\pi i} \left( cT^{\mathrm{m}}+c\partial\xi\eta+cT^{\phi}+bc\partial c+\eta e^{\phi}
G^{\mathrm{m}}-\eta\partial\eta e^{2\phi}b\right), \label{eq:ZJ}
\end{equation}
where the energy-momentum tensors $T^{\mathrm{m}}, T^{\phi}$ and the matter supercurrent 
$G^{\mathrm{m}}$ are 
\begin{eqnarray}
& &T^{\mathrm{m}}=-\frac{1}{4\ap}\partial X^{\mu}\partial X_{\mu}-\frac{1}{2}\psi^{\mu}
\partial\psi_{\mu}, \qquad T^{\phi}=-\frac{1}{2}\partial\phi\partial\phi -\partial^2
\phi, \phantom{QQQQQ} \label{eq:ZK} \\
& &  G^{\mathrm{m}}=\frac{i}{\sqrt{2\ap}}\partial X^{\mu}\psi_{\mu}. \nonumber
\end{eqnarray}
After lengthy calculations we reach 
\begin{eqnarray}
Q_B|A_+^{(0)}\rangle&=&\int\frac{d^{10}k}{(2\pi)^{10}}\biggl[
\partial^2cc\left(\sqrt{\frac{\ap}{2}}k^{\mu}A_{\mu}^1(k)-iv(k)-iw(k)\right) \nonumber \\
& &{}+\partial cc\partial X^{\mu}\left(i\sqrt{\frac{\ap}{2}}k^2A^1_{\mu}(k)+k_{\mu}v(k)\right)
 \nonumber \\
& &{}+\partial cc\psi^{\mu}\psi^{\nu}\left(-i\ap k^2\sqrt{\frac{\ap}{2}}F_{\mu\nu}(k)\right) 
 \nonumber \\
& &{}+\partial cc\partial\phi\left( i\ap k^2w(k)\right) \nonumber \\
& &{}+\eta e^{\phi} c\partial\psi^{\mu}\left(-A^1_{\mu}(k)+A^2_{\mu}(k)-\sqrt{2\ap}ik_{\mu}w(k)
 \right) \nonumber \\
& &{}+\eta e^{\phi}c\psi^{\nu}\partial X^{\mu}\left(-ik_{\nu}A^1_{\mu}(k)+ik_{\mu}A^2_{\nu}(k)
 -F_{\mu\nu}(k)+\frac{1}{\sqrt{2\ap}}\eta_{\mu\nu}w(k)\right) \nonumber \\
& &{}+\eta e^{\phi}c\psi^{\rho}\psi^{\mu}\psi^{\nu}\left(i\ap k_{[\rho}F_{\mu\nu]}(k)\right) \nonumber \\
& &{}+\partial\eta e^{\phi}c\psi^{\mu}\left(-A^1_{\mu}(k)+A^2_{\mu}(k)-2i\ap k^{\nu}F_{\mu\nu}(k)
 -\sqrt{2\ap}ik_{\mu}w(k)\right) \label{eq:ZO} \\
& &{}+\eta \partial e^{\phi} c\psi^{\mu}\left(-A^1_{\mu}(k)+A^2_{\mu}(k)-2i\ap k^{\nu}F_{\mu\nu}(k)
 -2\sqrt{2\ap}ik_{\mu}w(k)\right) \nonumber \\
& &{}+\eta e^{\phi}\partial c\psi^{\mu}\left(\ap k^2A^2_{\mu}(k)-\sqrt{2\ap}ik_{\mu}v(k)\right)
  \nonumber \\
& &{}+\partial^2\eta\eta e^{2\phi}\left(-\sqrt{\frac{\ap}{2}}k^{\mu}A^2_{\mu}(k)+iv(k)+2iw(k)\right)
 \label{eq:ZP} \\
& &{}+\partial\eta\eta\partial e^{2\phi}\left(-\sqrt{\frac{\ap}{2}}k^{\mu}A^2_{\mu}(k)+iv(k)+
 \frac{5}{2}iw(k)\right) \nonumber \\
& &{}+\partial\eta\eta e^{2\phi}\partial X^{\mu}\left(\frac{i}{\sqrt{2\ap}}(A^1_{\mu}(k)-
 A^2_{\mu}(k))\right) \label{eq:ZQ} \\
& &{}+\partial\eta\eta e^{2\phi}\psi^{\mu}\psi^{\nu}\left(\sqrt{2\ap}k_{[\nu}A^2_{\mu]}(k)
 +\frac{\sqrt{2\ap}}{2i}F_{\mu\nu}(k)\right) \label{eq:ZR} \\
& &{}+\partial\eta\eta e^{2\phi}bc(2iw(k))\biggr] e^{ikX}(0)|0\rangle, \label{eq:ZS}
\end{eqnarray}
where we have used the OPEs 
\begin{equation}
X^{\mu}(z)X^{\nu}(w)\sim -2\ap\eta^{\mu\nu}\log (z-w), \quad 
\psi^{\mu}(z)\psi^{\nu}(w)\sim\frac{\eta^{\mu\nu}}{z-w}, \label{eq:ZT}
\end{equation}
and eqs.(\ref{eq:ZH}). 
${}_{[\ldots]}$ denotes the antisymmetrization operation 
\begin{eqnarray*}
A_{[\mu}B_{\nu]}&=&\frac{1}{2!}(A_{\mu}B_{\nu}-A_{\nu}B_{\mu}), \\
A_{[\mu}B_{\nu}C_{\rho]}&=&\frac{1}{3!}(A_{\mu}B_{\nu}C_{\rho}+A_{\nu}B_{\rho}C_{\mu}
+A_{\rho}B_{\mu}C_{\nu}-A_{\mu}B_{\rho}C_{\nu}-A_{\nu}B_{\mu}C_{\rho}-A_{\rho}B_{\nu}C_{\mu}).
\end{eqnarray*}
The on-shell conditions are obtained from $Q_B|A_+^{(0)}\rangle =0$. 
The full set of resulting 15 equations, however, must be highly redundant 
because we have only five independent fields. First we choose as four independent equations 
the non-dynamical ones derived from the expressions~(\ref{eq:ZP})--(\ref{eq:ZS}), 
\begin{eqnarray}
& &w(k)=0,\quad A^1_{\mu}(k)=A^2_{\mu}(k), \label{eq:ZUb} \\
& &v(k)=-\sqrt{\frac{\ap}{2}}ik^{\mu}A^2_{\mu}(k), \label{eq:ZUa} \\
& &F_{\mu\nu}(k)=ik_{\mu}A^2_{\nu}(k)-ik_{\nu}A^2_{\mu}(k), \label{eq:ZU} 
\end{eqnarray}
by which the auxiliary fields $w,v,F_{\mu\nu},A^1_{\mu}$ can be eliminated. 
Then the fifth equation from~(\ref{eq:ZO}) implies the Maxwell equation 
\[ k^{\mu}F_{\mu\nu}(k)=0 \]
for the field strength tensor determined in~(\ref{eq:ZU}). One can see that the remaining ten equations 
are automatically satisfied. Among them is a Bianchi identity $ik_{[\rho}F_{\mu\nu]}=0$. 
\medskip

Let us consider the gauge degree of freedom. 
The gauge parameter $\Lambda$, which has ghost number 0 and picture number 0, 
has only one component $\lambda$ at the massless level, 
\begin{equation}
\Lambda=\int\frac{d^{10}k}{(2\pi)^{10}}i\sqrt{\frac{\ap}{2}}\lambda(k)e^{ikX}. \label{eq:ZV}
\end{equation}
At the linearized level, the gauge transformation law~(\ref{eq:gauge}) reduces to 
\begin{eqnarray}
\delta A_+^{(0)}=\frac{1}{\ap}Q_B\Lambda&=&\int\frac{d^{10}k}{(2\pi)^{10}}\biggl(
\frac{i}{\sqrt{2\ap}}ik_{\mu}\lambda(k)c\partial X^{\mu} \label{eq:ZW} \\ 
& &{}+ik_{\mu}\lambda(k)\eta e^{\phi}\psi^{\mu}+i\sqrt{\frac{\ap}{2}}k^2\lambda(k)\partial c
\biggr)e^{ikX}. \nonumber
\end{eqnarray}
Comparing it with the expansion~(\ref{eq:ZI}), we can read off the gauge transformation 
law for the component fields: 
\begin{eqnarray}
& &\delta A^1_{\mu}(k)=\delta A^2_{\mu}(k)=ik_{\mu}\lambda(k), \nonumber \\
& &\delta v(k)=\sqrt{\frac{\ap}{2}} k^2\lambda(k),\quad 
\delta F_{\mu\nu}(k)=\delta w(k)=0, \label{eq:ZX}
\end{eqnarray}
which are consistent with the equations of motion. In the Feynman-Siegel gauge
$b_0|A_+\rangle=0$, the coefficient $v$ of $\partial c$ is set to zero. 
Via the field equation~(\ref{eq:ZUa}), $v=0$ means $k^{\mu}A^2_{\mu}(k)=0$. 
Therefore, after eliminating the auxiliary fields using the linearized equations 
of motion, the Feynman-Siegel gauge condition 
implies the Lorentz gauge $k^{\mu}A^2_{\mu}(k)=0$ for the physical gauge field. 
\smallskip

Plugging (\ref{eq:ZUb})--(\ref{eq:ZU}) into (\ref{eq:ZI}), we get 
\begin{eqnarray}
A_+^{(0)}(z)&=&\int\frac{d^{10}k}{(2\pi)^{10}}\Biggl(\frac{i}{\sqrt{2\ap}}A^2_{\mu}(k)
c\partial X^{\mu}+A^2_{\mu}(k)\eta e^{\phi}\psi^{\mu} \label{eq:ZY} \\ 
& &{}+\sqrt{2\ap}k_{[\mu}A^2_{\nu]}(k)c\psi^{\mu}
\psi^{\nu}+\sqrt{\frac{\ap}{2}}k^{\mu}A^2_{\mu}(k)\partial c\Biggr) e^{ikX}(z). \nonumber
\end{eqnarray}
This `on-shell' vertex operator in fact  
coincides with the one obtained by acting with the picture-raising operator 
\begin{equation}
\mathrm{X}=c\partial \xi +e^{\phi}G^{\mathrm{m}}+e^{2\phi}b\partial\eta+\partial 
(e^{2\phi}b\eta) \label{eq:picturechange}
\end{equation}
on the massless vertex $V$ in the $-1$ picture, 
\begin{equation}
V(w)=\int\frac{d^{10}k}{(2\pi)^{10}}\Bigl(-\tilde{A}_{\mu}(k)ce^{-\phi}\psi^{\mu}e^{ikX}(w)
-i\tilde{v}(k)c\partial c\partial\xi e^{-2\phi}e^{ikX}(w)\Bigr). \label{eq:ZZ}
\end{equation}
Generically, $\displaystyle\lim_{z\to w}\mathrm{X}(z)V(w)$ contains a divergent piece: 
\begin{eqnarray}
\mathrm{X}(z)V(w)&=&\int\frac{d^{10}k}{(2\pi)^{10}}\biggl[\frac{1}{z-w}\left(\sqrt{2\ap}k^{\mu}\tilde{A}_{\mu}
(k)-2i\tilde{v}(k)\right)c(w)+\frac{i}{\sqrt{2\ap}}\tilde{A}_{\mu}(k)c\partial X^{\mu}(w) \nonumber \\
& &{}+\tilde{A}_{\mu}(k)\eta e^{\phi}\psi^{\mu}(w)+\sqrt{2\ap}k_{[\mu}\tilde{A}_{\nu]}(k)
c\psi^{\mu}\psi^{\nu}(w)+i\tilde{v}(k)\partial c(w) \label{eq:YA} \\ 
& &{}+\left(\sqrt{2\ap}k^{\mu}\tilde{A}_{\mu}(k)-2i\tilde{v}(k)\right)c\partial\phi(w)
+\cO (z-w)\biggr] e^{ikX(w)}. \nonumber
\end{eqnarray}
However, this divergent contribution can be removed by setting 
$\tilde{v}(k)=-\sqrt{\frac{\ap}{2}}ik^{\mu}\tilde{A}_{\mu}(k)$, 
which is one of the field equations obtained from the on-shell condition 
$Q_BV=0$. At the same time, the resulting expression agrees with~(\ref{eq:ZY}) 
if we identify $\tilde{A}_{\mu}$ with $A^2_{\mu}$. 
\medskip

Let us return to the computation of the action. 
Noting that $\partial X^{\mu}e^{ikX}, \partial c$ and $\partial \phi$ 
are no longer primary fields, we find 
\begin{eqnarray}
I\circ A^{(0)}_+(z)&=&\int\frac{d^{10}k}{(2\pi)^{10}}|z^{-2}|^{\ap k^2}\biggl[\frac{i}{\sqrt{2\ap}}
A^1_{\mu}(k)c\partial X^{\mu}+A^2_{\mu}(k)\eta e^{\phi}\psi^{\mu} \nonumber \\
& &{}+ \frac{\sqrt{2\ap}}{2i}F_{\mu\nu}(k)c\psi^{\mu}\psi^{\nu}+iv(k)\partial c 
+iw(k)c\partial\phi \label{eq:YB} \\
& &{}+z\left(-\sqrt{2\ap}k^{\mu}A^1_{\mu}(k)+2iv(k)+2iw(k)\right)c\biggr] e^{ikX}
\left(-\frac{1}{z}\right). \nonumber 
\end{eqnarray}
The final step is to substitute the expressions for $Q_BA_+^{(0)}(z)$ and $I\circ A^{(0)}_+(z)$ into 
\begin{equation} 
\llk Y_{-2}|A_+^{(0)},Q_BA_+^{(0)}\rrk=\lim_{z\to 0}\left\langle Y(i)Y(-i)
I\circ A^{(0)}_+(z)\ Q_BA_+^{(0)}(z)\right\rangle_{\mathrm{UHP}} \label{eq:YC}
\end{equation}
and evaluate numerous correlators using the OPEs~(\ref{eq:ZH}), (\ref{eq:ZT}) and 
the normalization~(\ref{eq:ZE}). The fully off-shell action for 
the massless component fields finally becomes  
\begin{eqnarray}
S^{(0)}&=&\frac{1}{2\ap g_o^2}\int\frac{d^{10}k}{(2\pi)^{10}}\biggl[-\eta^{\mu\nu}A^1_{\mu}(-k)
A^2_{\nu}(k)+\frac{1}{2}\eta^{\mu\nu}A^1_{\mu}(-k)A^1_{\nu}(k) \label{eq:YD} \\
& &{}+\frac{1}{2}\eta^{\mu\nu}A^2_{\mu}(-k)A^2_{\nu}(k)-2\ap \eta^{\mu\nu}ik^{\rho}A^2_{\mu}(-k)
F_{\nu\rho}(k)-\sqrt{2\ap}ik^{\mu}A^2_{\mu}(-k)w(k) \nonumber \\
& &{}+\frac{\ap}{2}F^{\mu\nu}(-k)F_{\mu\nu}(k)+\frac{5}{2}w(-k)w(k)+2v(-k)w(k)\biggr], \nonumber 
\end{eqnarray}
where $\eta^{\mu\nu}=\mathrm{diag}(-+\ldots +)$ is the spacetime metric. The set of 
five equations of motion derived by varying the action with respect to the field variables 
coincides with the previous one which has been found from $Q_BA_+^{(0)}=0$, 
as it should be.\footnote{It would not be the case if we had chosen $\cZ$ as the 
double-step inverse picture-changing operator~\cite{UZ}.}
Surprisingly, the above expression does not contain any contributions 
from the `Klein-Gordon operator' $c_0L_0^{\mathrm{m}}\sim\ap p^2$ in $Q_B$. 
From the (anomalous) $\phi$-charge conservation, one may na\"{\i}vely expect that 
$\llk Y_{-2}|A,\oint\frac{dz}{2\pi i}cT^{\mathrm{m}}A\rrk$ 
is non-vanishing if $A$ has $\phi$-charge $+1$. 
However, this correlator actually vanishes for the string field of the form 
$\eta e^{\phi}\cV(X^{\mu},\psi^{\mu})$ with $\cV$ denoting an arbitrary vertex operator 
made out of the matter fields, because 
$\langle\partial\xi(i)\partial\xi(-i)\eta ( -1/z ) \eta(z)\rangle_{\xi\eta}=0$ for any $z$. 
Nevertheless, the usual kinetic term for the physical gauge field $A_{\mu}^2$ can 
be obtained after integrating out the auxiliary fields by their equations of motion: 
\begin{eqnarray}
S^{(0)}[A^2]&=&\frac{1}{g_o^2}\int\frac{d^{10}k}{(2\pi)^{10}}\left(-\frac{1}{4}\cF^{\mu\nu}(-k)
\cF_{\mu\nu}(k)\right)\nonumber \\ &=&\frac{1}{g_o^2}\int d^{10}x\left(-\frac{1}{4}\cF^{\mu\nu}(x)
\cF_{\mu\nu}(x)\right), \label{eq:Maxwell}
\end{eqnarray}
where we have Fourier-transformed to the position space as 
$ \cF_{\mu\nu}(k)=\int d^{10}x \cF_{\mu\nu}(x)e^{-ikx},$ 
and $\cF_{\mu\nu}$ is the field strength tensor for the gauge potential, 
$ \cF_{\mu\nu}(x)=\partial_{\mu}A^2_{\nu}(x)-\partial_{\nu}A^2_{\mu}(x).$
Needless to say, the action~(\ref{eq:Maxwell}) is exactly the Maxwell action we are familiar with. 
Here, we can at last answer the question raised at the beginning of this section: 
Since the above kinetic term for the gauge field is accompanied by the standard 
coefficient $-\frac{1}{4}$, we conclude that 
the sign of the overall multiplicative factor in front of the string field theory action 
should be \textit{plus}, as already indicated in eq.(\ref{eq:ZA}), if we use the 
normalization convention~(\ref{eq:ZE}) of the correlator.

\sectiono{Including the GSO($-$) Tachyon}\label{sec:GSO-}

\subsection{Precise definition of the vertices}\label{sec:def}
The defining property of the GSO($-$) states is that they have odd world-sheet 
spinor numbers, where we assign to $\psi^{\mu},e^{q\phi}$ world-sheet 
spinor numbers 1 and $q$, respectively. If we restrict ourselves to the 
subspace of ghost number 1, it then follows that the GSO($-$) string field $A_-$ is 
Grassmann-\textit{even} and contains states of half-integer--valued conformal weights. 
First of all, since $A_-$ has different Grassmannality from the GSO($+$) string field $A_+$, 
it seems that they fail to obey common algebraic relations. 
This problem can be resolved by attaching the $2\times 2$ internal Chan-Paton matrices 
to the string fields and the operator insertions as~\cite{Berkovits,BSZ,ABG} 
\begin{eqnarray}
& &\widehat{Q}_B= Q_B\otimes\sigma_3,\quad \widehat{Y}_{-2}
=Y_{-2}\otimes \sigma_3, \nonumber \\
& &\widehat{A}=A_+\otimes\sigma_3+A_-\otimes i\sigma_2. \label{eq:YE}
\end{eqnarray}
Due to the fact that $A_-$ has half-integer weights $h_-$, $A_-$ changes its sign 
under the conformal transformation $\cR_{2\pi}$ representing the $2\pi$ rotation of the unit 
disk, namely 
\begin{equation}
\cR_{2\pi}\circ A_-(z)=(\cR_{2\pi}^{\prime}(z))^{h_-}A_-(\cR_{2\pi}(z))=e^{2\pi ih_-}A_-(z)
=-A_-(z). \label{eq:YF}
\end{equation}
This in particular means that an additional minus sign arises in the cyclicity relation, 
\begin{eqnarray}
& &\llk Y_{-2}|A_-,B_-\rrk=-\llk Y_{-2}|B_-,A_-\rrk, \label{eq:YG}\\
& &\llk Y_{-2}|A_-,B_1*B_2\rrk=-\llk Y_{-2}|B_1,B_2*A_-\rrk. 
\end{eqnarray}
Then, the cubic superstring field theory action including both NS($\pm$) string fields can be written 
as~\cite{ABKM1,ABG}
\begin{eqnarray}
S&=&\frac{1}{2g_o^2}\mathrm{Tr}\left[\frac{1}{2\ap}\llk \widehat{Y}_{-2}|\widehat{A},\widehat{Q}_B
\widehat{A}\rrk+\frac{1}{3}\llk \widehat{Y}_{-2}|\widehat{A},\widehat{A}*\widehat{A}\rrk\right] \nonumber \\
&=&\frac{1}{g_o^2}\Biggl[\frac{1}{2\ap}\llk Y_{-2}|A_+,Q_BA_+\rrk +\frac{1}{3}\llk Y_{-2}|A_+,
A_+*A_+\rrk \label{eq:YH} \\ & & \ \ {}+\frac{1}{2\ap}\llk Y_{-2}|A_-,Q_BA_-\rrk +
\llk Y_{-2}|A_-,A_+*A_-\rrk \Biggr]. \nonumber 
\end{eqnarray}
However, the last two terms still have sign ambiguities 
because of the square-roots in the conformal factors 
\[ (I^{\prime}(z))^{h_-}, \quad (f^{(3)\prime}_1(0))^{h_-}, \quad (f^{(3)\prime}_3(0))^{h_-}. \]
The authors of~\cite{BSZ} proposed a natural prescription to this problem in the case of 
the disk representation of the string vertices, and in addition showed how to translate it into the 
UHP representation: 
\begin{eqnarray}
& &\mbox{If the conformal maps $f^{(n)}_k$ defining the $n$-string vertex have the property that} \nonumber \\
& &\mbox{all $f^{(n)}_k(0)$ are real and satisfy 
$f^{(n)}_1(0)<f^{(n)}_2(0)<\ldots <f^{(n)}_n(0)$,} \label{eq:prescription} \\
& &\mbox{then we should choose the positive sign for all $(f^{(n)\prime}_k(0))^{1/2}$. } \nonumber 
\end{eqnarray}
In this paper we will follow this prescription and write down explicit expressions for 
the 2- and 3-string vertices. 
For the 3-string vertex, the prescription~(\ref{eq:prescription}) can immediately be applied because 
our definition~(\ref{eq:ZD}) of $f^{(3)}_k$ satisfies the condition 
\[ f^{(3)}_1(0)=-\sqrt{3}<f^{(3)}_2(0)=0<f^{(3)}_3(0)=\sqrt{3}. \] 
Hence we take 
\begin{equation}
(f^{(3)\prime}_1(0))^{h_-}=(f^{(3)\prime}_3(0))^{h_-}\equiv\bigg|\left(\frac{8}{3}\right)^{h_-}\bigg|. 
\label{eq:YI}
\end{equation}
In the case of the 2-string vertex, however, we have to be more careful. 
We define $\cR_{\theta}$ to be a conformal map 
corresponding to the rotation of the unit disk by an angle $\theta$, 
\begin{equation}
\cR_{\theta}(z)=h^{-1}\left(e^{i\theta}h(z)\right), \label{eq:YK}
\end{equation} 
which forms an Abelian subgroup of $SL(2,\aaru)$. 
Noting that the inversion can be expressed as 
$I(z)=h^{-1}(e^{-\pi i}h(z))=\cR_{-\pi}(z)$, we write the 2-vertex~(\ref{eq:ZB}) as 
\begin{equation}
\llk Y_{-2}|A,B\rrk=\left\langle Y(i)Y(-i)\ \cR_{-\pi}\circ 
A(0)\ B(0)\right\rangle_{\mathrm{UHP}}. \label{eq:YJ} 
\end{equation}
In order to make the above prescription applicable, 
we use the $SL(2,\aaru)$-invariance of the correlation function to
rewrite the 2-vertex in the following way, 
\begin{eqnarray}
\llk Y_{-2}|A,B\rrk&=&\lim_{\epsilon\to 0^+}\left\langle Y(i)Y(-i)\ \cR_{-\pi+2\epsilon}\circ 
A(0)\ \cR_{2\epsilon}\circ B(0)\right\rangle_{\mathrm{UHP}} \nonumber \\
& &\hspace{-2cm}=\lim_{\epsilon\to 0^+}(\cR^{\prime}_{2\epsilon}(0))^{h_A+h_B}(\cR^{\prime}_{-\pi}(z))^{h_A}
\left\langle Y(i)Y(-i) A(-1/z) B(z)\right\rangle_{\mathrm{UHP}}, \label{eq:YL} 
\end{eqnarray}
where we have defined $z\equiv \cR_{2\epsilon}(0)\simeq \epsilon \ (>0)$, and 
used the (de)composition law \linebreak $\cR_{-\pi+2\epsilon}=\cR_{-\pi}\circ\cR_{2\epsilon}$.
In addition, we have assumed $A,B$ to be primary fields for simplicity. Then, since 
$f^{(2)}_1(0)\equiv\cR_{-\pi+2\epsilon}(0)<0<f^{(2)}_2(0)\equiv\cR_{2\epsilon}(0)$, 
we can determine the prefactors of~(\ref{eq:YL}) to be 
\[(\cR^{\prime}_{2\epsilon}(0))^{h_A+h_B}(\cR^{\prime}_{-\pi}(z))^{h_A}=|(\sec\epsilon)^{2(h_A+h_B)}|
|z^{-2h_A}| \]
according to the prescription~(\ref{eq:prescription}). As for the first factor, 
nothing prevents us from taking the limit $\epsilon\to 0^+$ in advance, 
and it gives a factor of 1. Thus, we have finally found the 2-vertex to be given by 
\begin{equation}
\llk Y_{-2}|A,B\rrk=\lim_{z\to 0^+}\left\langle Y(i)Y(-i)\ \cR_{-\pi}\circ 
A(z)\ B(z)\right\rangle_{\mathrm{UHP}}, \label{eq:YM} 
\end{equation}
with the prescription for the conformal factor 
\begin{equation}
(\cR_{-\pi}^{\prime}(z))^h=z^{-2h}, \label{eq:YN}
\end{equation}
where we have used $|z|=z$ under $z\to 0^+$. Consistency of the composition law \linebreak
$\cR_{\pi}=\cR_{-\pi}\circ\cR_{2\pi}$ and the action of $\cR_{2\pi}$~(\ref{eq:YF}) 
then requires 
\begin{equation}
(\cR_{\pi}^{\prime}(z))^h=e^{2\pi ih}z^{-2h}. \label{eq:YO}
\end{equation}
For notational simplicity we shall use the conventional symbol $I$ as 
$I=\cR_{-\pi},\ I^{-1}=\cR_{\pi}$ with the prescriptions~(\ref{eq:YN}), (\ref{eq:YO}) included. 
That is to say, we \textit{define}
\begin{equation}
(I^{\prime}(z))^h=z^{-2h},\qquad ((I^{-1})^{\prime}(z))^h=e^{2\pi ih}z^{-2h}. \label{eq:inversion}
\end{equation}
Notice that $I^2\circ \Phi=\cR_{-2\pi}\circ\Phi=(-1)^{2h}\Phi$. 
Once we have found which of the square-root branches we should choose, 
we no longer need to specify how to take the limit $z\to 0$. 
To summarize, the 2-vertex is computed as 
\begin{equation}
\llk Y_{-2}|A,B\rrk =\lim_{z\to 0}\langle Y(i)Y(-i)I\circ A(z)\ B(z)\rangle_{\mathrm{UHP}}, 
\end{equation}
with the prescription~(\ref{eq:inversion}). 

So far we have had zero-momentum vertex operators in mind. 
We need some special care for the treatment of the momentum factor $e^{ikX}$, 
see Appendix~\ref{sec:AppB}. 

\subsection{Action for the tachyons}\label{sec:tachyon}
In the space of ghost number 1 and picture number 0, there are three 
negative-dimensional operators $c,\gamma,c\psi^{\mu}$. 
In this subsection 
we consider these `tachyon sectors', 
\begin{eqnarray}
|\widehat{A}\rangle&=&A_+^{(-1)}(0)|0\rangle\otimes\sigma_3+A_-^{(-1/2)}(0)|0\rangle\otimes i\sigma_2, 
\label{eq:AB} \\
A_+^{(-1)}(z)&=&\int\frac{d^{10}k}{(2\pi)^{10}}\sqrt{2}u(k)ce^{ikX}(z), \label{eq:AC} \\
A_-^{(-1/2)}(z)&=&\int\frac{d^{10}k}{(2\pi)^{10}}\left( t(k)\eta e^{\phi}+is_{\mu}(k)c\psi^{\mu}
\right) e^{ikX}(z). \label{eq:AD}
\end{eqnarray}
According to the prescriptions~(\ref{eq:inversion}) and (\ref{eq:ApT}), 
the inversion $I$ acts on $A_-^{(-1/2)}$ as 
\begin{equation}
I\circ A^{(-1/2)}_-(z)=\int\frac{d^{10}p}{(2\pi)^{10}}z|z^{-2}|^{\ap p^2}\left( t(p)\eta
e^{\phi}+is_{\mu}(p)c\psi^{\mu}\right)e^{ipX}\left(-\frac{1}{z}\right). \label{eq:YP}
\end{equation}
Plugging (\ref{eq:AC})--(\ref{eq:AD}) into the action~(\ref{eq:YH}), 
we get the component action for $u,t,s_{\mu}$ as 
\begin{eqnarray}
S&=&\frac{1}{g_o^2}\int\frac{d^{10}k}{(2\pi)^{10}}\frac{1}{2\ap}\biggl( u(-k)u(k)+\frac{1}{2}t(-k)t(k)
+\frac{1}{2}s^{\mu}(-k)s_{\mu}(k)+\sqrt{2\ap}ik_{\mu}s^{\mu}(-k)t(k)\biggr) \nonumber \\
& &{}+\frac{1}{g_o^2}\int\frac{d^{10}k_1d^{10}k_2d^{10}k_3}{(2\pi)^{20}}\delta^{10}(k_1+k_2+k_3)
\frac{9\sqrt{2}}{16}K^{-\ap (k_1^2+k_2^2+k_3^2)}t(k_1)u(k_2)t(k_3), \label{eq:YT}
\end{eqnarray}
where $K=3\sqrt{3}/4$. The standard kinetic term for the physical tachyon field $t$ is obtained 
only after eliminating the auxiliary field $s_{\mu}$ by its equation of motion 
\begin{equation}
s_{\mu}(k)+\sqrt{2\ap}ik_{\mu}t(k)=0. \label{eq:YU}
\end{equation}
Substituting (\ref{eq:YU}) back into (\ref{eq:YT}) and Fourier-transforming it, we obtain 
\begin{equation}
S=\frac{1}{g_o^2}\int d^{10}x\left[-\frac{1}{2}\left(-\frac{1}{\ap}\right) u(x)^2-\frac{1}{2}
(\partial_{\mu}t(x))^2-\frac{1}{2}\left(-\frac{1}{2\ap}\right)t(x)^2+\frac{9\sqrt{2}}{16}
\tilde{u}(x)\tilde{t}(x)^2\right], \label{eq:YV}
\end{equation}
where we have defined 
\[ \tilde{u}(x)=\exp \left(-\ap \ln \frac{4}{3\sqrt{3}}\partial^2\right)u(x), \quad 
\tilde{t}(x)=\exp \left(-\ap \ln \frac{4}{3\sqrt{3}}\partial^2\right)t(x). \]
Looking at the quadratic terms, we find that the physical tachyon field $t$ has correct 
kinetic and mass terms. On the other hand, 
the field $u$ lacks its kinetic term, so that it has non-dynamical equation of motion $u=0$ at 
the linearized level. Therefore, $u$ is indeed an auxiliary field and does not appear in the 
physical perturbative spectrum. Nevertheless $u$ can have significant effects on non-perturbative 
physics through the cubic interactions with other fields. 
\medskip

We conclude this section by noting that, if we substitute~(\ref{eq:YU}) into (\ref{eq:AD}), we 
again find that the resulting vertex operator 
\[ (\eta e^{\phi}+\sqrt{2\ap}ck_{\mu}\psi^{\mu})e^{ikX} \]
coincides with the one obtained by acting on the $-1$-picture vertex $-ce^{-\phi}e^{ikX}$ with 
the picture-raising operator $\mathrm{X}$~(\ref{eq:picturechange}). Hence, in order to analyse the fully 
off-shell dynamics of this theory we should use the intrinsically 0-picture vertex operators 
like (\ref{eq:ZI}) and (\ref{eq:AD}), instead of the picture-changed ones.

\sectiono{Reality Conditions}\label{sec:reality}
We go on to discuss the reality condition of the string field. As in the bosonic case, 
we represent it by combining the hermitian conjugation with the BPZ conjugation. 

\subsection{Preliminaries}\label{sec:preliminaries}
In terms of vertex operators, the BPZ conjugation is nothing but the conformal transformation by 
the inversion $I(z)=-1/z$. However, as discussed in the last section, its action on  
operators of half-integer weight contains sign ambiguity. Here we define the BPZ conjugation 
as $I$ with the prescription~(\ref{eq:inversion}). Then, its action on the $n$-th oscillator 
mode $\varphi_n$ of an arbitrary primary field $\varphi(z)$ becomes 
\begin{eqnarray}
\mathrm{bpz}(\varphi_n)&=&\mathrm{bpz}\left(\oint\frac{dz}{2\pi i}z^{n+h-1}\varphi(z)\right)
\equiv\oint\frac{dz}{2\pi i}z^{n+h-1}I\circ\varphi(z) \nonumber \\
&=&\oint\frac{dz}{2\pi i}z^{n-h-1}\varphi\left(-\frac{1}{z}\right)=(-1)^{-n+h}\varphi_{-n}, \label{eq:bpz}
\end{eqnarray}
which also holds for fields of half-integer weight (in which case $n$ takes half-integer values 
in the NS sector). BPZ conjugation is a linear map (\textit{i.e.} not accompanied 
by the complex conjugation), and preserves the order of operators. Since bpz 
satisfies $\mathrm{bpz}^2=(-1)^{2 h}$, the distinction between bpz and $\mathrm{bpz}^{-1}$ 
is important.\footnote{In the state formalism, bpz and hc are conventionally used to represent maps from a 
vector space $\cH$ to its dual space $\cH^*$, while bpz$^{-1}$ and hc$^{-1}$ from $\cH^*$ 
to $\cH$. In terms of vertex operators, there is no such distinction, so bpz and bpz$^{-1}$ 
differ only by~(\ref{eq:inversion}). }

Generically, for a field $\varphi(z)$ of conformal weight $h$ having the mode expansion 
$\varphi(z)=\sum_{n=-\infty}^{\infty}\varphi_n z^{-n-h}$, the hermitian conjugation, 
denoted by hc or $\dagger$, is taken as $(\varphi_n)^{\dagger}=\pm\varphi_{-n}$, together with 
the complex conjugation on $z$, $z^{\dagger}\to z^*$. Then we have 
\begin{equation}
(\varphi(z))^{\dagger}=\pm\sum_n\frac{\varphi_{-n}}{(z^*)^{n+h}}=
\pm (z^*)^{-2h}\varphi\left(\frac{1}{z^*}\right), \label{eq:herm}
\end{equation}
where the upper sign is for a 
\textit{hermitian} field and the lower sign for an \textit{antihermitian} field. 
This sign must be chosen so as not to contradict the commutation relations 
\begin{eqnarray}
& & [\alpha^{\mu}_n,\alpha_m^{\nu}]=n \eta^{\mu\nu}\delta_{n+m,0},\quad 
\{\psi^{\mu}_r,\psi^{\nu}_s\}=\eta^{\mu\nu}\delta_{r+s,0}, \label{eq:comm} \\
& & \{ c_n,b_m\} =\delta_{n+m,0}, \quad\qquad [\gamma_r, \beta_s]=\delta_{r+s,0}. \nonumber 
\end{eqnarray}
Noting that the hermitian conjugation reverses the order of operators as 
$(AB)^{\dagger}=B^{\dagger}A^{\dagger}$, we adopt  
\begin{eqnarray}
& & (\alpha^{\mu}_n)^{\dagger}=\alpha^{\mu}_{-n}, \quad (\psi^{\mu}_r)^{\dagger}=\psi^{\mu}_{-r}, \quad 
(b_n)^{\dagger}=b_{-n}, \quad (c_n)^{\dagger}=c_{-n}, \label{eq:dagger} \\
& &(\gamma_r)^{\dagger}=\gamma_{-r},\quad (\beta_r)^{\dagger}=-\beta_{-r}. \nonumber 
\end{eqnarray}
In other words, $\frac{i}{\sqrt{2\ap}}\partial X^{\mu}(z)=\sum_n\alpha_n^{\mu}z^{-n-1}, \psi^{\mu}, b,c$ 
and $\gamma$ are hermitian fields, while $\beta$ is an antihermitian field. 
The hermitian conjugation is an antilinear map in the sense that 
$(\lambda A)^{\dagger}=A^{\dagger}\lambda^*$ for $\lambda\in\shii$, and by definition it is idempotent, 
$(A^{\dagger})^{\dagger}=A$. Hence we do not distinguish hc from $\mathrm{hc}^{-1}$. 
\smallskip

In the following we will consider the composition map $\mathrm{bpz}\circ\mathrm{hc}$,
\begin{equation} 
\mathrm{bpz}\circ\mathrm{hc}(\varphi(z))=I\circ [(\varphi(z))^{\dagger}]
=\pm\varphi(-z^*). \label{eq:YX} 
\end{equation}
Notice that each $z$-derivative $\partial$ flips the hermiticity of fields: 
\begin{eqnarray}
(\partial\varphi(z))^{\dagger}&=&\left(\sum_n\frac{(-n-h)\varphi_n}{z^{n+h+1}}\right)^{\dagger}=\pm\sum_n
\frac{(-n-h)\varphi_{-n}}{(z^*)^{n+h+1}}=\pm\sum_n\frac{(n-h)\varphi_n}{(1/z^*)^{n+h+1}(z^*)^{2h+2}}
\nonumber \\ &=&\pm\left(-\frac{1}{(z^*)^{2(h+1)}}\sum_n\frac{(-n-h)\varphi_n}{(1/z^*)^{n+h+1}}
-\frac{2h}{(z^*)^{2h+1}}\sum_n\frac{\varphi_n}{(1/z^*)^{n+h}}\right) \nonumber \\
&=&\mp (z^*)^{-2(h+1)}\partial\varphi\left(\frac{1}{z^*}\right)\mp 2h(z^*)^{-2h-1}\varphi
\left(\frac{1}{z^*}\right), \label{eq:YY}
\end{eqnarray}
which shows that the hermiticity of $\partial\varphi$ is opposite to that of $\varphi$. 
The second term reflects the non-primary nature of $\partial\varphi$. In fact, in calculating 
the composition map $\mathrm{bpz}\circ\mathrm{hc}$ this extra term is precisely cancelled 
and we have 
\[ \mathrm{bpz}\circ\mathrm{hc}(\partial\varphi(z))=I\circ [(\partial\varphi(z))^{\dagger}]
=\mp\partial\varphi(-z^*). \] 

In order to discuss the reality condition in the fermionized language, we must reveal the hermiticity 
properties of the ghosts $\xi,\eta, e^{q\phi}$. From the abbreviated form of 
the fermionization formula we have 
\begin{equation}
\gamma^{\dagger}=(e^{\phi})^{\dagger}\eta^{\dagger}=-\eta^{\dagger}(e^{\phi})^{\dagger}, \qquad 
\beta^{\dagger}=(\partial\xi)^{\dagger}(e^{-\phi})^{\dagger}=-(e^{-\phi})^{\dagger}
(\partial\xi)^{\dagger}. \label{eq:YZ}
\end{equation}
If we require that they be consistent with the hermiticity properties of $\gamma,\beta$, namely 
\begin{equation}
\gamma(z)^{\dagger}=z^*\gamma(1/z^*) \quad \mbox{ and } \quad 
\beta(z)^{\dagger}=-(z^*)^{-3}\beta(1/z^*), \label{eq:XA}
\end{equation}
then it must be true that either $\eta$ or $e^{\phi}$ is antihermitian, and that 
both $e^{-\phi}$ and $\partial\xi$ are hermitian or antihermitian. Given that 
$e^{q\phi}=1+q\phi +\ldots$ contains the unit operator in it, one finds that $e^{q\phi}$ 
cannot be antihermitian. In addition, it follows from the result~(\ref{eq:YY}) that the hermiticity 
of $\partial\xi$ is opposite to that of $\xi$. All in all, we have found that:
\begin{eqnarray}
& &\eta \mbox{ is antihermitian} \quad :\quad  \eta_n^{\dagger}=-\eta_{-n}, \quad \eta(z)^{\dagger}=-(z^*)^{-2}
\eta(1/z^*), \nonumber \\
& &\xi \mbox{ is antihermitian} \quad :\quad  \xi_n^{\dagger}=-\xi_{-n}, \quad \xi(z)^{\dagger}=-\xi(1/z^*), 
\label{eq:XC} \\
& &:e^{q\phi}: \mbox{ (as well as } \phi) \mbox{ is hermitian} \quad :\quad  (:e^{q\phi(z)}:)^{\dagger}
=(z^*)^{q(q+2)}:e^{q\phi(1/z^*)}:. \nonumber
\end{eqnarray}
The rules (\ref{eq:XC}) are of course consistent with the commutation relation 
\[ \{\eta_m,\xi_n\}=\delta_{m+n,0}. \]

\subsection{Reality of the NS actions}\label{subsec:reality}
\underline{\textbf{modified cubic theory}} \vspace{2mm} \\
Making use of the tools prepared in the last subsection, we discuss what conditions on the string 
field guarantee the reality of the cubic action. We will show that for the NS($+$) sector 
in the 0-picture, whose structure is rather similar to that of bosonic string theory, the following 
condition on the vertex operator representation of the string field works well: 
\begin{equation}
\mathrm{bpz}\circ\mathrm{hc}(A_+(z))\equiv I\circ [A_+(z)^{\dagger}]=A_+(-z^*), \label{eq:XF}
\end{equation}
or equivalently, 
\begin{equation}
A_+(z)^{\dagger}=I^{-1}\circ A_+(-z^*). \label{eq:XG}
\end{equation}
(In the NS($+$) sector the distinction between $\mathrm{bpz}=I$ and $\mathrm{bpz}^{-1}=I^{-1}$ is 
irrelevant.) This in particular means that, by taking the limit $z\to 0$, 
$\mathrm{bpz}\circ\mathrm{hc}(|A_+\rangle)=|A_+\rangle$ in the state formalism. 
As mentioned in section~\ref{sec:massless}, the reality condition~(\ref{eq:XF}) on the string 
field $A_+^{(0)}$~(\ref{eq:ZI}) implies (\ref{eq:ZN}). Note that $\partial c$ and 
$\partial\phi$ are antihermitian. 

Next we consider the NS($-$) sector. From the form of the action~(\ref{eq:YH}) 
one immediately finds that the NS($-$) string field $A_-$ enters 
the action only quadratically. It then follows that not only ``real" string field but also 
``pure imaginary" string field gives rise to a real-valued action. We fix this ambiguity 
by requiring the real physical component fields to have the correct kinetic terms. 
Since we have obtained in the last section the correctly-looking kinetic term 
$-\frac{1}{2}(\partial_{\mu}t)^2$ for the physical tachyon field $t$ after eliminating the 
auxiliary vector $s_{\mu}$, we take this tachyon field to be real: $t(x)^*=t(x)$ or 
$t(k)^*=t(-k)$. Together with the vertex operator to which the tachyon is associated, it satisfies 
\begin{eqnarray}
& &\mathrm{bpz}\circ\mathrm{hc}\left(\int\frac{d^{10}k}{(2\pi)^{10}}t(k)(\eta e^{\phi}+
\sqrt{2\ap}ck_{\mu}\psi^{\mu})e^{ikX}(z)\right) \nonumber \\
& &\hspace{7mm} =\int\frac{d^{10}k}{(2\pi)^{10}} z^*|z^{-2}|^{\ap k^2}I\circ 
\left[e^{-ikX}(-e^{\phi}\eta+\sqrt{2\ap}k_{\mu}\psi^{\mu}c)\left(1/z^*\right)\right]
t(k)^* \nonumber \\
& &\hspace{7mm} =\int\frac{d^{10}k}{(2\pi)^{10}}t(-k)(\eta e^{\phi}+\sqrt{2\ap}c(-k_{\mu})\psi^{\mu})
e^{-ikX}(-z^*) \nonumber \\
& &\hspace{7mm} =\int\frac{d^{10}k}{(2\pi)^{10}}t(k)(\eta e^{\phi}+\sqrt{2\ap}ck_{\mu}\psi^{\mu})
e^{ikX}(-z^*), \label{eq:XH} 
\end{eqnarray}
where in the last line we have converted the integration variable from $k$ to $-k$. 
Extending it to the whole NS($-$) sector, we impose the following \textit{reality condition} 
on the NS($-$) string field, 
\begin{equation}
\mathrm{bpz}\circ\mathrm{hc}(A_-(z))\equiv I\circ [A_-(z)^{\dagger}]=A_-(-z^*), \label{eq:XI}
\end{equation}
where we must be careful to follow the sign convention~(\ref{eq:inversion}). 
\medskip 

Now we prove in the CFT language that the conditions~(\ref{eq:XF}), (\ref{eq:XI}) 
guarantee the reality of the cubic action~(\ref{eq:YH}).
First we note that the real string fields satisfy 
\begin{equation}
I\circ [(Q_BA_{\pm}(z))^{\dagger}]=-(-1)^{\mathrm{Grassmann}(A_{\pm})}
Q_BA_{\pm}(-z^*)=\pm Q_BA_{\pm}(-z^*), 
\label{eq:XJ}
\end{equation}
which can be shown from the hermiticity property of $Q_B$, or can be explicitly verified by 
looking at the expressions~(\ref{eq:ZO})--(\ref{eq:ZS}) for $Q_BA_+^{(0)}$ and $Q_BA_-^{(-1/2)}$. 
From~(\ref{eq:XF}) and (\ref{eq:XI}) it follows that 
\[ (I\circ A_{\pm}(z))^{\dagger}=\pm A_{\pm}(-z^*). \]
Then, the complex conjugate of the quadratic part of the action is calculated as 
\begin{eqnarray}
\llk Y_{-2}|A_{\pm},Q_BA_{\pm}\rrk^*&=&\lim_{z\to 0}\left\langle (Q_BA_{\pm}(z))^{\dagger}
(I\circ A_{\pm}(z))^{\dagger}(-Y(-i))(-Y(i))\right\rangle \nonumber \\
& &\hspace{-2.5cm}=(\pm)^2\lim_{z\to 0}\left\langle Y(i)Y(-i)A_{\pm}(-z^*)I^{-1}\circ (Q_BA_{\pm}
(-z^*))\right\rangle \nonumber \\
& &\hspace{-2.5cm}=\lim_{z^{\prime}\to 0}\left\langle Y(i)Y(-i) I\circ A_{\pm}(z^{\prime})
Q_BA_{\pm}(z^{\prime})\right\rangle=\llk Y_{-2}|A_{\pm},Q_BA_{\pm}\rrk, \label{eq:XK}
\end{eqnarray}
where we have used the facts that $Y(\pm i)$ is an antihermitian primary field of conformal weight 0, 
that the Grassmannality of $A_{\pm}$ and that of $Q_BA_{\pm}$ are different so that they commute with 
no sign factor, and that the correlator is invariant under the $SL(2,\aaru)$ transformation $I$. 
This shows that the quadratic part is indeed real. To examine the cubic term, we expand the string 
field as $A=\sum\Phi_i$, with each $\Phi_i$ having a definite conformal weight $h_i$.  
Then the cubic part of the action can be written as the sum of terms of the form 
\begin{eqnarray}
S_{(i,j,k)}&=&\llk Y_{-2}|\Phi_i,\Phi_j*\Phi_k\rrk +\llk Y_{-2}|\Phi_k,\Phi_j*\Phi_i\rrk \nonumber \\
&=&\left\langle Y(i)Y(-i)f^{(3)}_1\circ \Phi_i(0)f^{(3)}_2\circ\Phi_j(0)
f^{(3)}_3\circ\Phi_k(0)\right\rangle_{\mathrm{UHP}}+(i\leftrightarrow k). \label{eq:XL}
\end{eqnarray}
For simplicity we will assume $\Phi_i$'s to be primary, but the argument can be generalized 
to the non-primary case. Since the conformal factors $(f^{(3)\prime}_i(0))^h$ are real and 
$f^{(3)\prime}_1(0)=f^{(3)\prime}_3(0)=\frac{8}{3}$, we find 
\begin{eqnarray}
\hspace{-0cm}
S_{(i,j,k)}^*&=&(f^{(3)\prime}_1(0))^{h_i}(f^{(3)\prime}_2(0))^{h_j}(f^{(3)\prime}_3(0))^{h_k}
\nonumber \\ & &{}\times
\left\langle \Phi_k(\sqrt{3})^{\dagger}\Phi_j(0)^{\dagger}\Phi_i(-\sqrt{3})^{\dagger}Y(-i)Y(i)
\right\rangle_{\mathrm{UHP}}+(i\leftrightarrow k) \nonumber \\
&=&(f^{(3)\prime}_3(0))^{h_i}(f^{(3)\prime}_2(0))^{h_j}(f^{(3)\prime}_1(0))^{h_k} \nonumber \\
& &{}\times
\left\langle Y(i)Y(-i)I^{-1}\circ\Phi_k(-\sqrt{3})I^{-1}\circ\Phi_j(0)I^{-1}\circ\Phi_i(\sqrt{3})
\right\rangle_{\mathrm{UHP}}+(i\leftrightarrow k) \nonumber \\
&=&\left\langle Y(i)Y(-i)f^{(3)}_1\circ\Phi_k(0)f^{(3)}_2\circ\Phi_j(0)f^{(3)}_3\circ
\Phi_i(0)\right\rangle_{\mathrm{UHP}}+(i\leftrightarrow k), \label{eq:XM}
\end{eqnarray}
where we have used the reality conditions~(\ref{eq:XG}), (\ref{eq:XI}) and the $I$-invariance 
of the correlator. The last expression of (\ref{eq:XM}) is equal to $S_{(i,j,k)}$ thanks to 
the presence of the $(i\leftrightarrow k)$ term. 
This completes the proof of the reality of the modified NS cubic action. 
Incidentally, the fact that the ordering of the operators 
has been reversed after taking the complex conjugation should be related to the orientation reversal 
appearing in the functional form of the reality condition
$ \Phi[X^{\mu}(\sigma)]^*=\Phi[X^{\mu}(\pi-\sigma)]$. 
\bigskip  

\noindent
\underline{\textbf{Witten's cubic theory}} \vspace{2mm} \\
In Witten's original proposal for open superstring field theory~\cite{Witten2} we propose 
the following reality conditions for the NS($\pm$) string fields $V_{\pm}$ in 
the $-1$-picture, 
\begin{equation}
I\circ \left(V_{\pm}(z)^{\dagger}\right)=-V_{\pm}(-z^*). \label{eq:XN}
\end{equation}
The $-$ sign originates from the fact that the picture-changing operators $\mathrm{X}, Y$ are antihermitian. 
The easiest way to show how the above conditions work would be to demonstrate some examples: 
The vertex operators to which the tachyon and the massless gauge field are associated are 
$ce^{-\phi}e^{ikX}\ (\sim -Y\cdot \gamma e^{ikX})$ and $\psi^{\mu}ce^{-\phi}e^{ikX}$ respectively, 
and they satisfy 
\begin{eqnarray*}
I\circ\left[\left(\int\frac{d^{10}k}{(2\pi)^{10}}t(k)ce^{-\phi}e^{ikX}(z)\right)^{\dagger}\right]
&=&-\int\frac{d^{10}k}{(2\pi)^{10}}t(k)^*ce^{-\phi}e^{-ikX}(-z^*), \\
I\circ\left[\left(\int\frac{d^{10}k}{(2\pi)^{10}}A_{\mu}(k)\psi^{\mu}ce^{-\phi}e^{ikX}(z)\right)^{\dagger}
\right]&=&-\int\frac{d^{10}k}{(2\pi)^{10}}A_{\mu}(k)^*\psi^{\mu}ce^{-\phi}e^{-ikX}(-z^*),
\end{eqnarray*}
so the conditions~(\ref{eq:XN}) lead to $t(k)^*=t(-k)$ and $A_{\mu}(k)^*=A_{\mu}(-k)$. 
The cubic action is given by~\cite{Witten2,0004112,Raeymaekers} 
\begin{eqnarray}
S&=&\frac{1}{g_o^2}\biggl[\frac{1}{2\ap}\langle V_+,Q_BV_+\rangle+\frac{1}{3}\llk \mathrm{X}|V_+,V_+*V_+
\rrk \nonumber \\ & & \ \ +\frac{1}{2\ap}\langle V_-,Q_BV_-\rangle+\llk \mathrm{X}|V_-,V_+*V_-\rrk \biggr], 
\label{eq:XO}
\end{eqnarray}
where the 2-string vertex $\langle\cdots , \cdots\rangle$ is defined by the simple BPZ inner product, 
\[ \langle A,B\rangle =\lim_{z\to 0}\langle I\circ A(z) \ B(z)\rangle_{\mathrm{UHP}} \]
with the sign prescription~(\ref{eq:inversion}), while the cubic interaction vertex is defined as 
\[ \llk \mathrm{X}|A,B*C\rrk=\left\langle \mathrm{X}(i)\ f^{(3)}_1\circ A(0) f^{(3)}_2\circ B(0) f^{(3)}_3\circ C(0)
\right\rangle_{\mathrm{UHP}}, \]
with (\ref{eq:YI}). The proof of the reality of the action is almost identical to the modified cubic 
case: For the quadratic terms, considering that $Y(i)Y(-i)$ played no special r\^{o}le in the previous 
proof, the same argument as in~(\ref{eq:XK}) holds true. For the cubic terms, we need to use 
$I\circ\mathrm{X}(i)=\mathrm{X}(-\frac{1}{i}=i)$ and $\mathrm{X}(i)^{\dagger}=-\mathrm{X}(1/i^*=i)$. 
This $-$ sign cancels the extra three $-$ signs arising from~(\ref{eq:XN}), giving rise to 
the real action. 
\bigskip

\noindent
\underline{\textbf{Berkovits' non-polynomial theory}} \vspace{2mm} \\
The string fields $\Phi_{\pm}$ in this theory have vanishing ghost and picture numbers, and in a 
partial gauge $\xi_0\Phi_{\pm}=0$, $\Phi_{\pm}$ are in a one-to-one correspondence to the above 
$-1$-picture string fields $V_{\pm}$ through $\Phi_{\pm}=:\xi V_{\pm}:$~\cite{sP}. Therefore the 
reality conditions on $\Phi_{\pm}$ can be deduced from those on $V_{\pm}$ as 
\begin{equation}
I\circ \left(\Phi_{\pm}(z)^{\dagger}\right)=\mp\Phi_{\pm}(-z^*), \label{eq:XP}
\end{equation}
because $\xi$ is antihermitian and Grassmann-odd. 
The WZW-like action is given by 
\begin{eqnarray}
S&=&\frac{1}{4g_o^2}\mathrm{Tr}\bllk\left(e^{-\widehat{\Phi}}\widehat{Q}_B
e^{\widehat{\Phi}}\right)\left(e^{-\widehat{\Phi}}\widehat{\eta}_0
e^{\widehat{\Phi}}\right) \nonumber \\
& &{}-\int_0^1dt\left(e^{-t\widehat{\Phi}}\partial_te^{t\widehat{\Phi}}\right)
\left\{\left(e^{-t\widehat{\Phi}}\widehat{Q}_Be^{t\widehat{\Phi}}\right),
\left(e^{-t\widehat{\Phi}}\widehat{\eta}_0e^{t\widehat{\Phi}}\right)\right\}
\brrk \label{eq:WZW1} \\ &=&\frac{1}{2g_o^2}\sum_{M,N=0}^{\infty}\frac{(-1)^N}{(M+N+2)!}
\left({M+N \atop N}\right)\mathrm{Tr}\bllk\left(\widehat{Q}_B\widehat{\Phi}\right)
\widehat{\Phi}^M\left(\widehat{\eta}_0\widehat{\Phi}\right)\widehat{\Phi}^N
\brrk, \label{eq:WZW2}
\end{eqnarray}
where in the last line we have expanded the exponentials in a formal power series. 
We will now give the proof of the reality of the action~(\ref{eq:WZW2}) 
in the GSO-projected case.  
The GSO($-$) sector can be incorporated with a little more care. 

The action~(\ref{eq:WZW2}) can be arranged in the form 
\begin{eqnarray}
S&=&\frac{1}{2g_o^2}\sum_{M,N=0}^{\infty}\frac{1}{(M+N+2)!}\biggl[(-1)^N
\left({M+N \atop N}\right)\llk(Q_B\Phi)\Phi^M(\eta_0\Phi)\Phi^N\rrk \nonumber \\
& &\hspace{4.1cm} {}+(-1)^M\left({N+M \atop M}\right)\llk(Q_B\Phi)\Phi^N(\eta_0\Phi)\Phi^M
\rrk\biggr] \nonumber \\
&=&\frac{1}{2g_o^2}\sum_{M,N=0}^{\infty}\frac{(-1)^N}{(M+N+2)!}\left({M+N \atop N}\right)
\label{eq:XQ} \\
& &{}\times\biggl[\llk(Q_B\Phi)\Phi^M(\eta_0\Phi)\Phi^N\rrk
+(-1)^{M-N+1}\llk\Phi^N(\eta_0\Phi)\Phi^M(Q_B\Phi)\rrk\biggr], \nonumber 
\end{eqnarray}
where we have used the cyclicity of the bracket.\footnote{For more details about this theory, 
see the original papers~\cite{sP,BSZ} and reviews~\cite{0105230,0102085,DeSmet}. } 
Note that the factor of $1/2$ has been compensated for by taking the trace. 
Upon expanding the string field $\Phi=\sum\Phi_i$ as above, we find that, in order to prove the 
reality of the full action~(\ref{eq:XQ}), it is sufficient 
to show that the specific combination $\cL_1+(-1)^{M+N+1}\cL_2$ is real, where 
\begin{eqnarray}
& &\cL_1=\llk(Q_B\Phi_1)\Phi_2\cdots \Phi_{M+1}(\eta_0\Phi_{M+2})\Phi_{M+3}\cdots 
\Phi_{M+N+2}\rrk, \label{eq:XR} \\
& &\cL_2=\llk\Phi_{M+N+2}\cdots \Phi_{M+3}(\eta_0\Phi_{M+2})\Phi_{M+1}\cdots 
\Phi_{2}(Q_B\Phi_1)\rrk. \nonumber 
\end{eqnarray}
From the reality condition for the GSO($+$) string field~(the upper sign of (\ref{eq:XP})) 
it follows that 
\begin{equation}
(\eta_0\Phi(z))^{\dagger}=-I\circ (\eta_0\Phi(-z^*)), \qquad 
(Q_B\Phi(z))^{\dagger}=I\circ (Q_B\Phi(-z^*)). \label{eq:XS}
\end{equation}
If we further assume that $Q_B\Phi_1$ and $\eta_0\Phi_{M+2}$ are primary, 
then the complex conjugate of $\cL_1$ is calculated as 
\begin{eqnarray}
\cL_1^*&=&\Bigl\langle \tilde{f}^{(M+N+2)}_1\circ (Q_B\Phi_1(0))\tilde{f}^{(M+N+2)}_2\circ\Phi_2(0)\cdots 
\tilde{f}^{(M+N+2)}_{M+1}\circ\Phi_{M+1}(0) \label{eq:XT} \\ & &{}\times \tilde{f}^{(M+N+2)}_{M+2}
\circ(\eta_0\Phi_{M+2}(0))\tilde{f}^{(M+N+2)}_{M+3}\circ
\Phi_{M+3}(0)\cdots \tilde{f}^{(M+N+2)}_{M+N+2}\circ\Phi_{M+N+2}(0)\Bigr\rangle^*_{\mathrm{disk}} \nonumber \\
&=&\prod_{k=1}^{M+N+2}\left[ \tilde{f}^{(M+N+2)\prime}_k(0)^{h_k}\right]^*(-1)^{M+N+1}\Bigl\langle I\circ
\Phi_{M+N+2}(-z_{M+N+2}^*)\cdots I\circ\Phi_{M+3}(-z_{M+3}^*) \nonumber \\
& &{}\times I\circ\left(\eta_0\Phi_{M+2}(-z_{M+2}^*)\right)I\circ\Phi_{M+1}(-z_{M+1}^*)\cdots
I\circ\Phi_2(-z_2^*)I\circ\left( Q_B\Phi_1(-z_1^*)\right)\Bigr\rangle_{\mathrm{disk}}, \nonumber
\end{eqnarray} 
where the conformal maps and the values of their derivatives at the origin are defined as 
\begin{equation}
\tilde{f}^{(N)}_k(z)=e^{2\pi i\frac{2k-1}{2N}}\left(\frac{1+iz}{1-iz}\right)^{\frac{2}{N}}, \qquad 
\tilde{f}^{(N)\prime}_k(0)^h=\bigg|\left(\frac{4}{N}\right)^h\bigg|e^{2\pi ih(\frac{2k-1}{2N}+\frac{1}{4})},
\label{eq:XU}
\end{equation}
and $z_k\equiv \tilde{f}^{(M+N+2)}_k(0)=e^{2\pi i\frac{2k-1}{2(M+N+2)}}$. For later convenience, 
we have adopted different phase factors for $\tilde{f}^{(N)}_k(z)$ than~\cite{BSZ}. 
The complex conjugate of the conformal factor becomes 
\begin{eqnarray}
\left[ \tilde{f}^{(M+N+2)\prime}_k(0)^{h_k}\right]^*&=&\bigg|\left(\frac{4}{M+N+2}\right)^{h_k}\bigg|
\exp\left(-2\pi ih_k\left(\frac{2k-1}{2(M+N+2)}+\frac{1}{4}\right)\right) \nonumber \\
& &\hspace{-3cm}=\bigg|\left(\frac{4}{M+N+2}\right)^{h_k}\bigg|
\exp\left(2\pi ih_k\left(\frac{2(M+N+3-k)-1}{2(M+N+2)}+\frac{1}{4}\right)\right)e^{-2\pi ih_k(1+\frac{1}{2})}
\nonumber \\
& &\hspace{-3cm}=\tilde{f}^{(M+N+2)\prime}_{M+N+3-k}(0)^{h_k}e^{-3\pi ih_k}. \label{eq:XV}
\end{eqnarray}
Plugging it into~(\ref{eq:XT}) and performing the $SL(2,\aaru)$ transformations 
$I^{-1}$ and $z\to e^{\pi i}z$ inside the disk correlator, we find 
\begin{eqnarray}
\cL_1^*&=&\left(\prod_{k=1}^{M+N+2}\tilde{f}^{(M+N+2)\prime}_{M+N+3-k}(0)^{h_k}\right)
e^{-3\pi i\sum_{k=1}^{M+N+2}h_k}(-1)^{M+N+1}e^{\pi i\sum_{k=1}^{M+N+2}h_k} \label{eq:XW} \\
& &\hspace{-1cm}\times
\Bigl\langle \Phi_{M+N+2}(z_{M+N+2}^*)\cdots \Phi_{M+3}(z_{M+3}^*)\eta_0\Phi_{M+2}(z_{M+2}^*)
\Phi_{M+1}(z_{M+1}^*)\cdots\Phi_2(z_2^*)Q_B\Phi_1(z_1^*)\Bigr\rangle_{\mathrm{disk}}. \nonumber
\end{eqnarray}
The two phase factors cancel each other because the sum of the weights is always an integer. 
Using  
\begin{equation}
z_k^*=\tilde{f}^{(M+N+2)}_k(0)^*=e^{-2\pi i\frac{2k-1}{2(M+N+2)}}=e^{2\pi i\left(\frac{2(M+N+3-k)-1}{2(M+N+2)}
-1\right)}=z_{M+N+3-k}, \label{eq:XX}
\end{equation}
$\cL_1^*$ can further be rewritten as 
\begin{eqnarray}
\cL_1^*&=&(-1)^{M+N+1}\Bigl\langle \tilde{f}^{(M+N+2)}_1\circ\Phi_{M+N+2}(0)\cdots \tilde{f}^{(M+N+2)}_N\circ 
\Phi_{M+3}(0)\tilde{f}^{(M+N+2)}_{N+1}\circ (\eta_0\Phi_{M+2}(0)) \nonumber \\
& &{}\times \tilde{f}^{(M+N+2)}_{N+2}\circ\Phi_{M+1}(0)\cdots \tilde{f}^{(M+N+2)}_{M+N+1}\circ\Phi_2(0)
\tilde{f}^{(M+N+2)}_{M+N+2}\circ (Q_B\Phi_1(0))\Bigr\rangle_{\mathrm{disk}}, \nonumber 
\end{eqnarray}
which precisely coincides with $(-1)^{M+N+1}\cL_2$. This shows that 
$\cL_1+(-1)^{M+N+1}\cL_2$ is real.

\sectiono{Application to Tachyon Condensation}\label{sec:tachyonpotential}
\subsection{Homogeneous tachyon condensation on a non-BPS D-brane}
\begin{figure}[htbp]
	\begin{center}
	\scalebox{0.9}[0.9]{\includegraphics{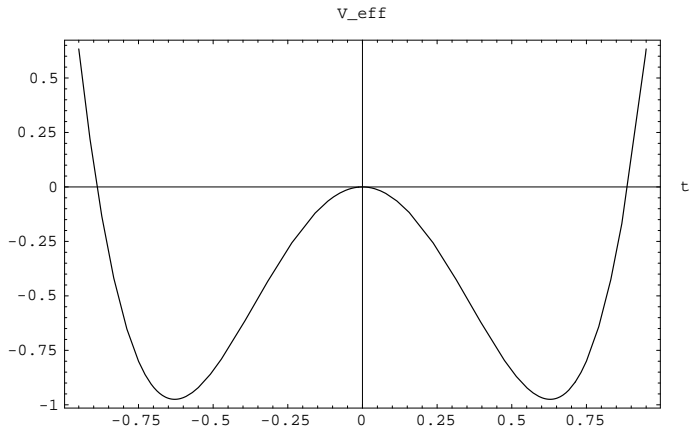}}
	\end{center}
	\caption{The effective tachyon potential at level $(\frac{1}{2},1)$.}
	\label{fig:potential}
\end{figure}

In this subsection we reconsider the problem of the static and spatially homogeneous 
tachyon condensation on a non-BPS D9-brane in the framework of level-truncated modified 
cubic superstring field theory, which was first investigated by Aref'eva, Belov, Koshelev and 
Medvedev~\cite{ABKM1} and further by Raeymaekers~\cite{Raeymaekers}. Its physical interpretation is, 
of course, the decay of the unstable D-brane. We assign to each component field $\phi_i$ the level 
number defined by $h_i+1$, where $h_i$ is the conformal weight 
of the vertex operator to which $\phi_i$ is associated, in such a way that the state of the lowest 
weight has level 0. Since the physical tachyon field $t$ we want to investigate is at level $1/2$ 
by this definition, we should start with the level $(1/2,1)$ approximation instead of 
$(0,0)$.\footnote{As usual, the `level $(N,M)$ truncation' means that the string field contains only terms 
of level less than or equal to $N$, and the action contains interaction terms of level less than or 
equal to $M$, where the level of an interaction term is defined to be the sum of the level numbers 
of the fields involved in it.} Let us first recall the mechanism of how the expected tachyon potential 
of the double-well form can be reproduced from the cubic action~(\ref{eq:YH}). The level $(1/2,1)$-truncated 
tachyon potential $V^{(\frac{1}{2},1)}$ can immediately be obtained by setting $u(x)$ and $t(x)$ to 
constants in~(\ref{eq:YV}), 
\begin{equation}
V^{(\frac{1}{2},1)}\equiv -\frac{S^{(\frac{1}{2},1)}}{V_{10}}=\frac{1}{g_o^2}
\left(-\frac{1}{2\ap}u^2-\frac{1}{4\ap}t^2-\frac{9\sqrt{2}}{16}u t^2\right). \label{eq:WA}
\end{equation}
To obtain the effective potential for $t$ we integrate out the auxiliary field $u$ 
at the tree-level, \textit{i.e.} by its equation of motion 
\begin{equation}
u=-\frac{9\sqrt{2}}{16}\ap t^2. \label{eq:WB}
\end{equation}
The resulting effective tachyon potential becomes~\cite{ABKM1} 
\begin{equation}
V^{(\frac{1}{2},1)}_{\mathrm{eff}}=\frac{1}{g_o^2}\left(-\frac{1}{4\ap}t^2+\frac{81}{256}\ap t^4
\right), \label{eq:WC}
\end{equation}
which is quartic and really takes the double-well form (see Fig.\ref{fig:potential}). 
In short, the tachyon potential with a qualitatively desirable profile has been obtained by integrating 
out an auxiliary field which sits at the level lower than the tachyon, despite the 
absence of genuinely higher order interactions in the action~(\ref{eq:YH}). 
This is in sharp contrast to the case of Berkovits' superstring field theory where 
the tachyon is the field of the lowest level and reproduces the quartic potential 
by itself~\cite{Berkovits}. In order to compare the depth of the 
potential with the D-brane tension quantitatively, we need a formula relating the open 
string coupling $g_o$ to the non-BPS D9-brane tension $\tilde{\tau}_9$. By applying the method 
invented in~\cite{Univ} we have found 
\begin{equation}
\tilde{\tau}_9=\frac{1}{2\pi^2g_o^2\ap{}^3} \label{eq:WD}
\end{equation}
in our convention. Then, the minimum value 
of the effective potential~(\ref{eq:WC}) can easily be evaluated as 
\begin{equation}
V_{\mathrm{eff}}^{(\frac{1}{2},1)}\big|_{\mathrm{min}}= -\frac{8}{81}\pi^2 \tilde{\tau}_9
\simeq -0.975\ \tilde{\tau}_9 \qquad \mathrm{at} \quad t=\pm \frac{4\sqrt{2}}{9\ap}\simeq\pm 
\frac{0.629}{\ap}. \label{eq:WE}
\end{equation}
According to the Sen's conjecture, the value of the tachyon potential at the minimum should cancel 
the tension of the unstable D-brane, so 
$V_{\mathrm{eff}}^{(\mathrm{exact})}\big|_{\mathrm{min}}=-\tilde{\tau}_9$. 
Hence we have found that about 97.5\% of the expected value has already been reproduced 
at the lowest level of approximation. This behavior of the minimum value of the potential 
is again very different from the case of Berkovits' theory, 
where only 61.7\% of the brane tension is obtained at the lowest level 
and the vacuum value gradually approaches $-\tilde{\tau}_9$ as the level is increased~\cite{DeSmet}.  
\medskip

As shown in~\cite{ABKM1}, the modified cubic action~(\ref{eq:YH}) is invariant under the 
$\zetto_2$ twist transformation $A_{\pm}\to \Omega A_{\pm}$, where $\Omega$ acts on 
each $L_0^{\mathrm{tot}}$-eigenstate as 
\begin{equation}
\Omega (\Phi)=\left\{
	\begin{array}{lcl}
	(-1)^{h_{\Phi}+1}\Phi & \mathrm{for} & \mbox{NS($+$) states} \quad (h_{\Phi}\in\zetto) \\
	(-1)^{h_{\Phi}+\frac{1}{2}}\Phi & \mathrm{for} & \mbox{NS($-$) states} \quad 
	\left(h_{\Phi}\in\zetto+\frac{1}{2}\right)
	\end{array}
\right. .\label{eq:WF}
\end{equation}
Due to this twist symmetry, all the twist-odd fields (\textit{e.g.} fields at levels $1,\frac{3}{2}$) 
can be set to zero without contradicting the equations of motion. (Note that the tachyon $t$ 
and the auxiliary scalar $u$ are twist-even.) 
Therefore we should include the level-2 fields at the next step. 

At level 2, we have 9 independent component fields in the so-called universal basis, 
\begin{eqnarray}
A^{(1)}_+&=&v_1\partial^2c+v_2 cT^{\mathrm{m}}+v_3 c:\partial\xi\eta:+v_4 cT^{\phi}
+v_5c\partial^2\phi \label{eq:WG} \\
& &{}+v_6\eta e^{\phi}G^{\mathrm{m}}+v_7:bc\partial c:+v_8\partial c\partial\phi+
v_9b\eta\partial\eta e^{2\phi}, \nonumber
\end{eqnarray}
where we are keeping the field $v_9b\eta\partial\eta e^{2\phi}$ of $\phi$-charge 2, which 
was dropped in~\cite{ABKM1}. 
Note that the reality condition~(\ref{eq:XF}) requires the component fields $v_i$ to be real. 
Substituting $A_+=\sqrt{2} u c+A_+^{(1)}$ and 
$A_-=t \eta e^{\phi}$ into (\ref{eq:YH}), we have computed the tachyon potential up to level 
(2,6), whose explicit expression is shown in Appendix~\ref{sec:AppA}. 
At this level, however, there are gauge degrees of freedom 
\begin{equation}
\Lambda_+^{(1)}=\lambda_1:bc:+\lambda_2\partial\phi. \label{eq:WGauge}
\end{equation}
In the following we will try several gauge-fixing conditions. 
\medskip

\begin{figure}[htbp]
	\begin{center}
	\scalebox{1.0}[1.0]{\includegraphics{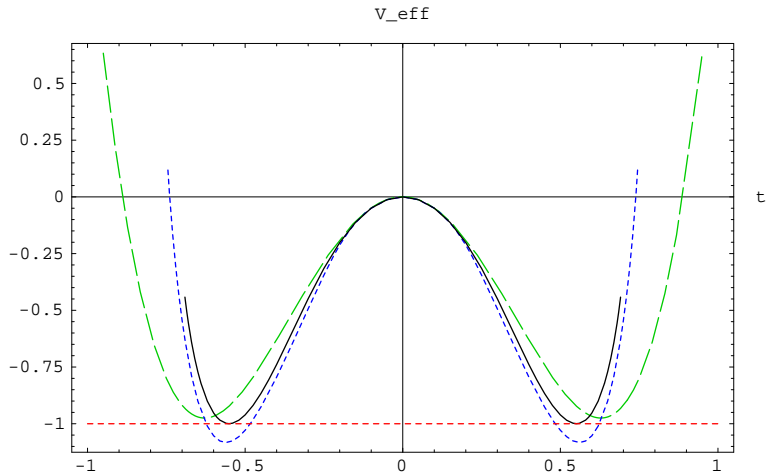}}
	\end{center}
	\caption{The effective tachyon potential in the Feynman-Siegel gauge 
	at level $(\frac{1}{2},1)$ (dashed line), 
	level (2,4) (dotted line) and level (2,6) (solid line). The dashed straight line 
	indicates the expected depth of $-1$. At level (2,6) the branch 
	ends at $t\simeq \pm 0.691$.}
	\label{fig:FS}
\end{figure}

\noindent
\underline{\textbf{The Feynman-Siegel gauge $b_0A_{\pm}=0$}}\\ 
First we choose the Feynman-Siegel gauge 
\begin{equation}
b_0 A_{\pm}(0)\equiv \oint\frac{dz}{2\pi i}zb(z)A_{\pm}(0)=0, \label{eq:WH}
\end{equation}
which implies $v_7=v_8=0$ at level 2. Its perturbative validity can be shown 
in the same way as in bosonic string field theory~\cite{SZ}. 
By extremizing the action~(\ref{eq:ApA}) under the conditions $v_7=v_8=0$ we can numerically 
look for the tachyon vacuum solution and calculate the depth of the potential. 
The results are: 
\[ V^{(2,4)}\big|_{\mathrm{min}}=-1.08273\ \tilde{\tau}_9, \qquad 
V^{(2,6)}\big|_{\mathrm{min}}=-0.999584\ \tilde{\tau}_9. \]
We have also calculated the effective tachyon potential at each level, whose profile 
is shown in Fig.\ref{fig:FS}.
The minimum value calculated at level (2,6) is surprisingly close to the expected value of $-1$ 
times the D9-brane tension, but we consider it as just a coincidence because it is not clear at all
even whether the minimum value of the potential is really 
converging or not. 
\smallskip

\begin{figure}[htbp]
	\begin{center}
	\scalebox{0.9}[0.9]{\includegraphics{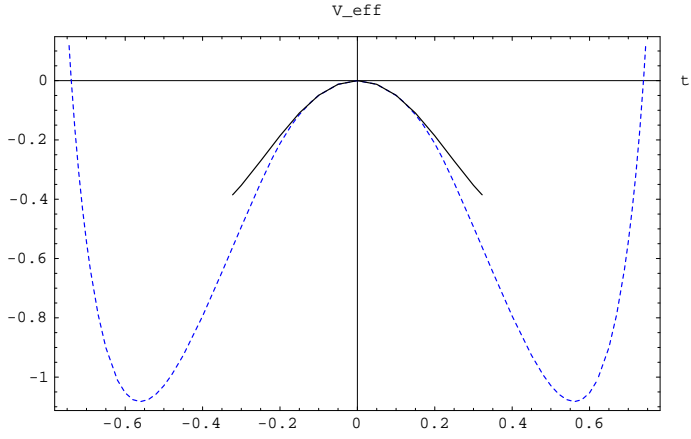}}
	\end{center}
	\caption{The effective tachyon potential in the Feynman-Siegel gauge 
	at level $(\frac{5}{2},5)$ (solid line). The branch ends at $t\simeq \pm 0.322$. 
	The dotted line shows the potential at level (2,4). }
	\label{fig:sing}
\end{figure}

The multiscalar tachyon potential at level $(\frac{5}{2},5)$ in the Feynman-Siegel gauge has been 
calculated by Raeymaekers~\cite{Raeymaekers}. 
He argued that, although there exists a candidate tachyon vacuum solution, 
the branch of the potential on which the candidate 
tachyon vacuum exists does not cross the unstable perturbative vacuum ($V_{\mathrm{eff}}(t=0)=0$), 
so that it should not be considered as the correct tachyon vacuum solution. In fact, 
when we used his multi-scalar lagrangian to calculate the effective tachyon potential 
starting from the perturbative vacuum, 
we have found that the branch connected to the perturbative vacuum 
hits a singularity \textit{before} it reaches a 
minimum~(see Fig.\ref{fig:sing}). 
In view of the result~\cite{ET} obtained in bosonic string field theory, 
it may indicate that the Feynman-Siegel gauge choice 
is no longer valid beyond this singularity. If this is the case, 
we have to find a good gauge choice which works well 
at level ($\frac{5}{2},5$) or higher. 
\medskip

\noindent
\underline{\textbf{$3v_2-3v_4+2v_5=0$ with $v_9=0$}}\\ 
In \cite{ABKM1} Aref'eva, Belov, Koshelev and Medvedev proposed the gauge choice 
$3v_2-3 v_4+2 v_5=0$ such that the terms linear in $v_6$ should vanish, and 
in a subsequent paper~\cite{ABKM2} they studied the validity of this gauge 
in the level truncation scheme. With this choice, the equations of motion admit a solution 
with $v_6=0$, which makes the analysis much simplified. They also proposed 
that the string field configurations should be restricted to the space of $\phi$-charge 
0 or 1. That is, if we expand the NS string field as $A=\sum_{q\in\zetto}A_q$ according to 
$\phi$-charge $q$, then we should set $A_q=0$ for $q\neq 0,1$.\footnote{Since 
the present author does not agree with this proposal, we do not make this restriction anywhere 
else in this paper.} This means that the coefficient $v_9$ of $b\eta\partial\eta e^{2\phi}$ 
is set to zero. We refer to the conditions $3v_2-3v_4+2v_5=0, v_9=0$ as `ABKM gauge' below. 
As already claimed in~\cite{ABKM1}, the solutions at levels (2,4) and (2,6) coincide with 
each other in this gauge, and we find
\[ V^{(2,4)}\big|_{\mathrm{min}}=V^{(2,6)}\big|_{\mathrm{min}}=-1.05474\ \tilde{\tau}_9, \]
which confirms their result.\footnote{Although the minimum value was 
reported to be 105.8\% in~\cite{ABKM1}, it should simply be a typo because we are using the same 
lagrangian as theirs (see Appendix \ref{sec:AppA}).}
\medskip

\noindent
\underline{$(b_1+b_{-1})A_{\pm}=0$}\\
In a pioneering paper \cite{PTY} Preitschopf, Thorn and Yost proposed a gauge choice (which 
we call `PTY gauge') 
\begin{equation}
(b_1+b_{-1})A_{\pm}(0)=\oint\frac{dz}{2\pi i}(1+z^2)b(z)A_{\pm}(0)=0, \label{eq:WI}
\end{equation}
and showed that the correct tree-level scattering amplitudes were obtained in this gauge. 
We have also used this gauge to look for the non-perturbative tachyon vacuum solution. The condition~(\ref{eq:WI}) 
relates the coefficient of the state $c_{-1}|\psi\rangle$ to that of $c_1|\psi\rangle$, 
where $|\psi\rangle$ is an arbitrary state of ghost number 0 and picture number 0 which contains 
neither $c_1$ nor $c_{-1}$. Up to level 2, only one state $c_{-1}|0\rangle\simeq\frac{1}{2}\partial^2c(0)$ 
contains the $c_{-1}$ mode, so the gauge condition~(\ref{eq:WI}) implies 
\[ v_1=-\frac{1}{\sqrt{2}}u. \]
With this condition, however, we have not found any suitable solution for the tachyon vacuum. 
For example, at level (2,4) we have found a solution with vevs $u\simeq -0.446$ and $t\simeq 0.553$, 
but its energy density is about 203\% of the expected value. At level (2,6), we have found no solution 
around the above point in the field configuration space. This indicates that the PTY gauge 
may not be useful in searching for the non-perturbative tachyon vacuum solution. 
\medskip

\noindent 
\underline{\textbf{Without gauge fixing}}\\

\begin{figure}[htbp]
	\begin{center}
	\scalebox{1.0}[1.0]{\includegraphics{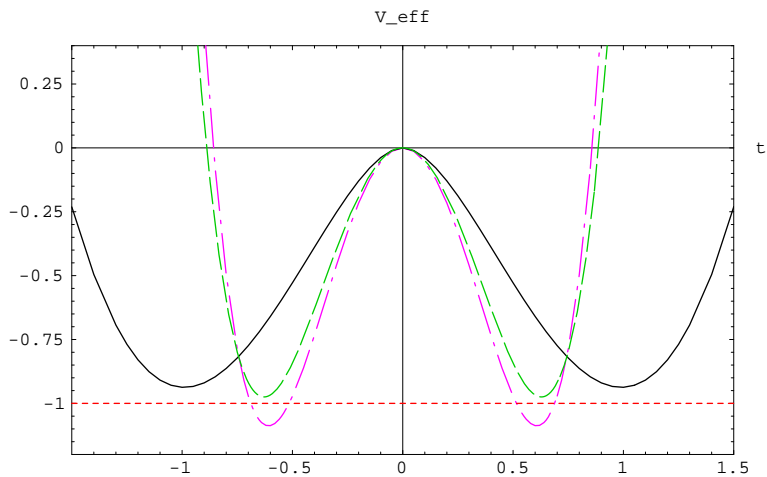}}
	\end{center}
	\caption{The gauge-unfixed effective tachyon potential at level $(\frac{1}{2},1)$ (dashed line), 
	level (2,4) (chain line) and level (2,6) (solid line). }
	\label{fig:unfix}
\end{figure}

Finally we look for the tachyon vacuum solution without any gauge-fixing conditions. 
From Fig.~\ref{fig:unfix} one sees that, at level (2,4), the effective tachyon potential in this case 
shows a similar behavior to the Feynman-Siegel gauge potential (Fig.~\ref{fig:FS}). 
Its depth is about 109\% of the expected D-brane tension. 
At the next level (2,6), however, the value of the tachyon field at the minimum becomes too large, 
although the potential depth $\simeq -0.937\ \tilde{\tau}_9$ may seem to be 
reasonable. 
Hence it is doubtful whether the effective tachyon potential without gauge-fixing 
really converges or not. 
\medskip

The results obtained in this subsection are summarized in Table~\ref{tab:potent}.
\begin{table}[htbp]
	\begin{center}
	\scalebox{0.95}[0.95]{
	\begin{tabular}{|c||c|c|c|c|}
	\hline
	level & Feynman-Siegel gauge & ABKM gauge & PTY gauge & gauge unfixed \\
	\hline\hline
	($\frac{1}{2},1$) & \multicolumn{4}{|c|}{$-0.974776$} \\
	\hline
	(2,4) & $-1.08273$ & $-1.05474$ & $-2.02738$ & $-1.08791$ \\
	\cline{1-2}\cline{4-5}
	(2,6) & $-0.999584$ &  & --- & $-0.937313$  \\
	\hline
	\end{tabular}}
	\end{center}
	\caption{The depth of the tachyon potential calculated in several gauges (normalized 
	by the non-BPS D9-brane tension).}
	\label{tab:potent}
\end{table}

\subsection{Non-perturbative vacuum on a BPS D-brane?}\label{subsec:ssb}
Given that there exists a negative-dimensional operator $c$ in the GSO($+$) sector, 
one might wonder whether it induces a `tachyon condensation' even in the GSO-projected theory, 
\textit{i.e.} on a \textit{BPS} D-brane. More than a decade ago, 
Aref'eva, Medvedev and Zubarev used modified cubic superstring field theory with 
the picture-changing operator $\cZ$~(\ref{eq:AMZpicturechange}) to explore such a possibility~\cite{AMZssb}. 
In this theory, the cubic self-interaction $u^3$ among the auxiliary field $u$ 
does not vanish, so that the effective potential for $u$ takes the `cubic form' 
just like the tachyon potential in bosonic string field theory. 
Then it becomes possible for $u$ to condense to the local minimum of its potential, 
though to our present knowledge we cannot give any physical interpretation to 
such a solution.  
They also argued that the spacetime supersymmetry was spontaneously broken in this vacuum. 
\medskip

\begin{figure}[htbp]
	\begin{center}
	\scalebox{1.3}[1.3]{\includegraphics{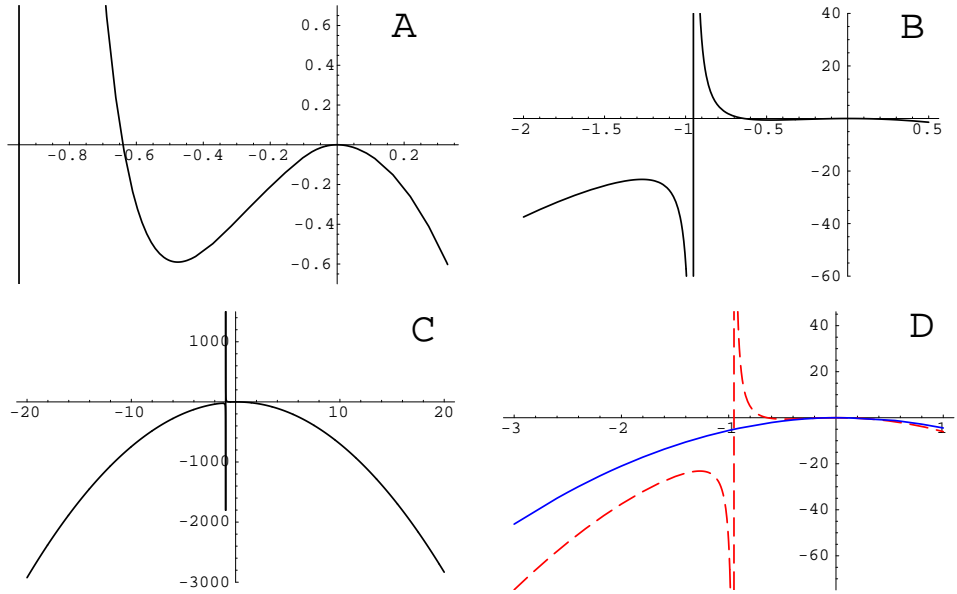}}
	\end{center}
	\caption{The Feynman-Siegel gauge effective potential for $u$ in the GSO-projected theory. A--C: 
	The level (2,4) potential at various ranges. D: The potential at level (2,4) (dashed line) 
	and at level (2,6) (solid line) where the singular structure has been resolved.}
	\label{fig:Vproj1}
\end{figure}

\begin{figure}[htbp]
	\begin{center}
	\scalebox{1.3}[1.3]{\includegraphics{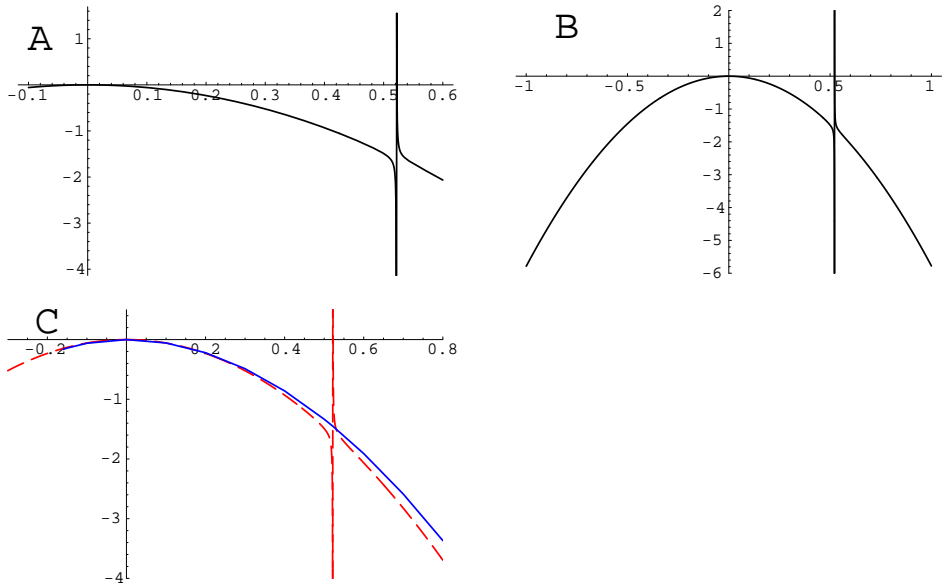}}
	\end{center}
	\caption{The gauge-unfixed effective potential for $u$ in the GSO-projected theory. A--B: 
	The level (2,4) potential at different ranges. C: The potential at level (2,4) (dashed line) 
	and at level (2,6) (solid line).}
	\label{fig:Vproj2}
\end{figure}

What happens if we carry out the same analysis in modified cubic superstring field theory 
with $Y_{-2}=Y(i)Y(-i)$, which is of our interest? The GSO-projected action can be obtained 
simply by setting all the GSO($-$) components to zero in the non--GSO-projected 
action~(\ref{eq:ApA}). At level (2,4) in the Feynman-Siegel gauge, the effective potential for $u$ 
seems to have a minimum at $u\simeq -0.476$ (Fig.~\ref{fig:Vproj1}A). 
However, this critical point, together with the singularity at $u\simeq -0.952$, disappears 
in the level (2,6) potential (Fig.~\ref{fig:Vproj1}D). Furthermore, without gauge fixing, 
there is no extremum in the effective potential up to level (2,6) 
(Fig.~\ref{fig:Vproj2}). 
From these results, we conclude that 
there are no locally stable vacua to which the auxiliary field $u$ condenses. 
This is in agreement with the expectation that the BPS D-brane is stable. 

\subsection{A brief survey of spatially inhomogeneous condensation}
An efficient method for constructing lower-dimensional D-branes as tachyon lump solutions 
in bosonic string field theory was invented by Moeller, Sen and Zwiebach~\cite{MSZ}, and 
it was shown in~\cite{0104230} that this method can also be applied to the case of 
Berkovits' superstring field theory where a kink solution on a non-BPS D-brane 
represents a BPS D-brane of one lower dimension. In this method 
we suppose that not only the oscillator non-zero modes
but also the center-of-mass momentum $e^{ikX}$ contributes 
to the level. For example, $ce^{ikX}$ and $\psi^{\mu}\eta e^{\phi}e^{ikX}$ 
have level numbers $\ap k^2$ and $\ap k^2+1$, respectively. 
The truncation of the string field at level $N$ means that 
we drop all terms in the string field with levels higher than $N$. 
Let us consider the field configurations which depend only on one spatial direction, 
say $x\equiv x^9$, and set $k_{\mu}=0$ for all $\mu\neq 9$. If we compactify the 
$x$-direction on a circle of radius $R$, the momentum $k\equiv k_9$ is discretized 
as $k_n=n/R$. As a result, the total number of degrees of freedom 
remains finite at any finite level 
even after the inclusion of the non-zero momentum modes. 
The computational framework based on the above procedure 
is called `modified level truncation scheme'.
\smallskip

\begin{figure}[htbp]
	\begin{center}
	\scalebox{1.6}[1.6]{\includegraphics{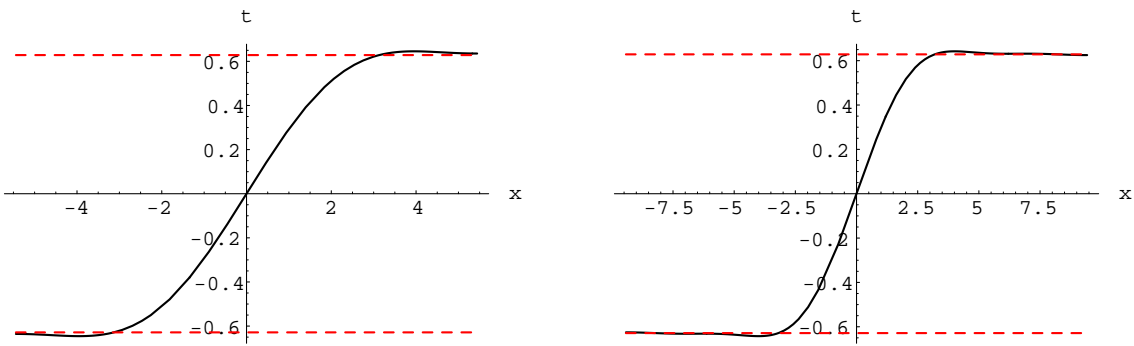}}
	\end{center}
	\caption{Kink solutions at $R=\sqrt{3\ap}$ (left) and at $R=3\sqrt{\ap}$ (right).}
	\label{fig:kink}
\end{figure}

Here we apply the above method to the modified cubic superstring field theory 
defined on a non-BPS D9-brane. By substituting 
\begin{equation}
u(x)=u_0+2 \sum_{n=1}^{n_{\mathrm{max}}}u_n\cos\frac{n}{R}x, \qquad 
t(x)=\sum_{r=1/2}^{r_{\mathrm{max}}}\tau_r\sin\frac{n}{R}x \label{eq:KB}
\end{equation}
into the action~(\ref{eq:YV}) and extremizing it with respect to $\{u_n,\tau_r\}$, 
we can find a solution which corresponds to the BPS D8-brane at the lowest 
level of approximation. More details are found in~\cite{0104230}. 
We show the two sets of results: level $(\frac{4}{3},\frac{19}{6})$ 
for $R=\sqrt{3\ap}$ and level $(\frac{67}{36},\frac{25}{6})$ for $R=3\sqrt{\ap}$. 
From the tachyon profile $t(x)$ shown in Fig.~\ref{fig:kink}, 
we see that the tachyon field correctly approaches one of the tachyon vacua 
in the asymptotic regions.  
The energy density $\cT_8$ of the kink solution relative to the BPS D8-brane tension $\tau_8$ 
can be calculated by the formula~\cite{0104230}
\begin{equation}
r\equiv\frac{\cT_8}{\tau_8}=\sqrt{2}\frac{R}{\sqrt{\ap}}(f(A_{\mathrm{kink}})-f(A_0)), 
\end{equation}
where $f=-S/(2\pi RV_9\tilde{\tau}_9)$, and $A_{\mathrm{kink}},A_0$ denote 
the kink solution and the tachyon vacuum solution, respectively. 
The expected value of $r$ is, of course, 1. We have found
\begin{eqnarray}
r=1.01499 & \mathrm{for} & R=\sqrt{3\ap}, \label{eq:ratios} \\
r=1.01441 & \mathrm{for} & R=3\sqrt{\ap}. \nonumber
\end{eqnarray}
Although we again regard these close agreements as accidental, these results suggest 
that the modified cubic theory truncated to low levels captures the 
quantitative as well as qualitative features of the space-dependent tachyon condensation. 
It would also be interesting to calculate the energy distribution of the kink solution 
in the $x$-direction, as was done in~\cite{Yang} for the lump solution 
in bosonic string field theory. 
\medskip

From the definition of the modified level it is clear that the modified level 
truncation scheme cannot be applied to the study of time-dependent solutions, 
because the level number is not bounded below if we allow  
large time-like momenta $k^2<0$, by which the level truncation procedure itself is invalidated. 
Instead, using the \textit{oscillator-level} truncation scheme (\textit{i.e.} the action~(\ref{eq:YV}))
Aref'eva \textit{et al.} found numerically a time-dependent solution of 
cubic superstring field theory equations of motion 
in which the tachyon starts rolling from the unstable vacuum and approaches one of the 
tachyon vacua in the asymptotic future~\cite{AJKV}. 
On the other hand, in bosonic string theory where the tachyon potential has its minimum 
at a finite distance away from the origin,  
nobody has succeeded so far in constructing a time-dependent solution with a desirable rolling profile 
(see \textit{e.g.}~\cite{MZ,Kluson,FH,MoeSch}).

\sectiono{Level Truncation Analysis in Vacuum Superstring Field Theory}\label{sec:VSFTsolutions}
In bosonic VSFT, Gaiotto, Rastelli, Sen and Zwiebach showed by the level truncation 
analysis that there exists a spacetime-independent solution 
whose form, up to an overall normalization, converges to the twisted butterfly state~\cite{GRSZ}. 
It is believed that this solution corresponds to a spacetime-filling D25-brane. 
This result 
can be considered as a piece of evidence for  
the usefulness of the level truncation calculations in VSFT. 
Here we will try a similar analysis in vacuum superstring field theory. 

\subsection{Cubic vacuum superstring field theory}\label{subsec:cubVSFT}
\begin{figure}[htbp]
	\begin{center}
	\scalebox{0.8}[0.8]{\includegraphics{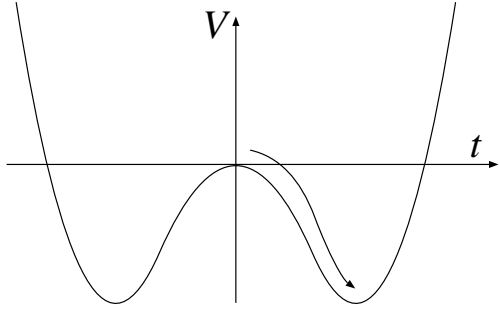}}
	\end{center}
	\caption{The reflection symmetry should be spontaneously broken after the 
	tachyon condensation.}
	\label{fig:z2break}
\end{figure}

Following an earlier work~\cite{ABG}, we proposed the following form of $\widehat{\cQ}$ 
as a candidate kinetic operator\footnote{The relative sign between $\Qev^{\mathrm{GSO}(+)}$ 
and $\Qev^{\mathrm{GSO}(-)}$, 
which is fixed by requiring that $\widehat{\cQ}$ satisfies the hermiticity, is different from that 
of ref.\cite{0208009} because we are obeying different sign conventions for the 2-string vertex: 
$(I^{\prime}(z))^h=z^{-2h}$ here while $(I^{\prime}(z))^h=e^{2\pi ih}z^{-2h}$ there. }
of vacuum superstring field theory~\cite{0208009}:
\begin{eqnarray}
& &\hspace{-5mm}\widehat{\cQ}=\Qod\otimes\sigma_3+\Qev\otimes (-i\sigma_2); \label{eq:VSFTop} \\
& &\Qod=\frac{1}{2i\varepsilon_r^2}(c(i)-c(-i))-\frac{q_1^2}{2}\oint\frac{dz}{2\pi i}
b\gamma^2(z), \nonumber \\
& &\Qev^{\mathrm{GSO}(+)}=\frac{q_1}{2i\varepsilon_r}(\gamma(i)-\gamma(-i)), \nonumber \\
& &\Qev^{\mathrm{GSO}(-)}=-\frac{q_1}{2\varepsilon_r}(\gamma(i)+\gamma(-i)), \nonumber 
\end{eqnarray}
with $\varepsilon_r\to 0$ and $q_1$ is some unknown constant. 
This operator was constructed such that:
\begin{itemize}
\item $\widehat{\cQ}$ should satisfy the axioms such as nilpotency, derivation property 
and hermiticity in order for $\widehat{\cQ}$ to be used to construct 
(classically) gauge invariant actions, 
\item $\widehat{\cQ}$ should have vanishing cohomology 
in order to support no perturbative physical open string 
degrees of freedom around the tachyon vacuum, 
\item $\widehat{\cQ}$ should preserve the twist symmetry of the action, 
\item $\Qev$ should be non-zero in order that 
the VSFT action does not possess the $\zetto_2$ GSO symmetry under $A_-\to -A_-$, 
because such a symmetry should be spontaneously broken 
after the tachyon condenses to one of the stable vacua (Fig.\ref{fig:z2break}).
\end{itemize}
In spite of some efforts~\cite{NSgs,0204138,Lechtenfeld,0208009}, 
no exact solution representing the unstable D9-brane has been found so far. 
\medskip

Cubic vacuum superstring field theory action is given by 
\begin{eqnarray}
S_V&=&\frac{\kappa_0}{2}\mathrm{Tr}\left[ \frac{1}{2}\llk \widehat{Y}_{-2}|\widehat{\cA},\widehat{\cQ}
\widehat{\cA}\rrk+\frac{1}{3}\llk \widehat{Y}_{-2}|\widehat{\cA},\widehat{\cA}*\widehat{\cA}\rrk
\right] \label{eq:VA} \\
&=&\kappa_0\biggl[\frac{1}{2}\llk Y_{-2}|\cA_+,\Qod \cA_+\rrk+\frac{1}{2}\llk Y_{-2}|\cA_-,
\Qod\cA_-\rrk+\llk Y_{-2}|\cA_-,\Qev^{\mathrm{GSO}(+)}\cA_+\rrk \nonumber \\
& &{}+\frac{1}{3}\llk Y_{-2}|\cA_+,\cA_+*\cA_+\rrk +\llk Y_{-2}|\cA_-,\cA_+*\cA_-\rrk\biggr], 
\nonumber
\end{eqnarray}
where we have set $\ap =1$ and $\kappa_0$ is some positive constant. 
Here, a surprising thing happens: Since $c(\pm i)$ are in the 
kernel of $Y(i)Y(-i)$, such terms in $\Qod$ give no contributions to the action. 
On the other hand, $\gamma(\pm i)$ in $\Qev$ are still non-vanishing because 
\[ \lim_{z\to i}Y(z)\eta e^{\phi}(i)=-ce^{-\phi}(i) \]
is finite. One may consider it is absurd that the $c$-ghost insertions at the 
open string midpoint vanish, but we will proceed anyway. 
After the rescaling $\cA_{\pm}\to \frac{q_1^2}{2}\cA_{\pm}$ of the string fields, 
the VSFT action can be arranged as 
\begin{eqnarray}
S_V&=&\kappa_0\left(\frac{q_1^2}{2}\right)^3
\biggl[\frac{1}{2}\llk Y_{-2}|\cA_+,Q_2 \cA_+\rrk+\frac{1}{2}\llk Y_{-2}|\cA_-,
Q_2\cA_-\rrk \nonumber \\ & &\hspace{1.8cm}{}+\frac{1}{i\epsilon}
\llk Y_{-2}|\cA_-,(\gamma(i)-\gamma(-i))\cA_+\rrk \label{eq:rescale} \\
& &\hspace{1.8cm}{}+\frac{1}{3}\llk Y_{-2}|\cA_+,\cA_+*\cA_+\rrk +\llk Y_{-2}|\cA_-,\cA_+*\cA_-\rrk\biggr], 
\nonumber
\end{eqnarray}
where $Q_2\equiv -\oint\frac{dz}{2\pi i}b\gamma^2(z)$ and $\epsilon\equiv q_1\varepsilon_r$. 
Inserting the expansion 
\begin{eqnarray}
\cA_+&=&\sqrt{2}uc+v_1\partial^2c+v_2 cT^{\mathrm{m}}+v_3 c:\partial\xi\eta:+v_4 cT^{\phi}
+v_5c\partial^2\phi \nonumber \\
& &{}+v_6\eta e^{\phi}G^{\mathrm{m}}+v_7:bc\partial c:+v_8\partial c\partial\phi+
v_9b\eta\partial\eta e^{2\phi}, \label{eq:VB} \\
\cA_-&=&t\eta e^{\phi} , \nonumber 
\end{eqnarray}
into (\ref{eq:rescale}), we obtain the action truncated up to level (2,6). 
Explicit expression of it is shown in Appendix~\ref{sec:AppA}. 
\smallskip

\begin{figure}[htbp]
	\begin{center}
	\scalebox{0.9}[0.9]{\includegraphics{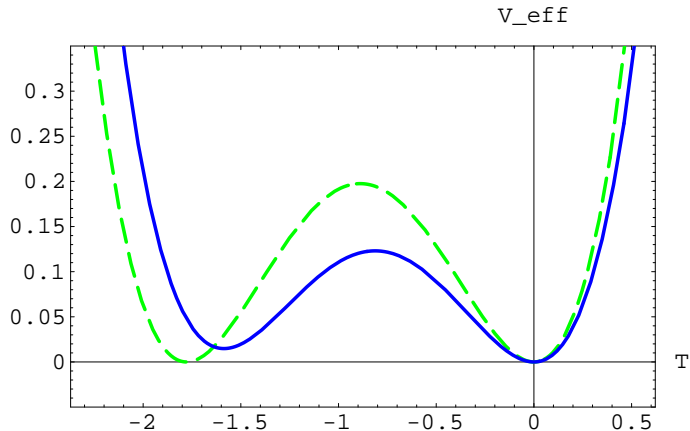}}
	\end{center}
	\caption{The cubic VSFT effective potential at level (1/2,1) (dashed line) 
	and at level (2,4) in the Siegel gauge (solid line). The horizontal 
	axis represents $T=\epsilon t$, while the vertical axis $\epsilon^4 V_{\mathrm{eff}}$.}
	\label{fig:VSFTpot1}
\end{figure}

\begin{figure}[htbp]
	\begin{center}
	\scalebox{0.8}[0.8]{\includegraphics{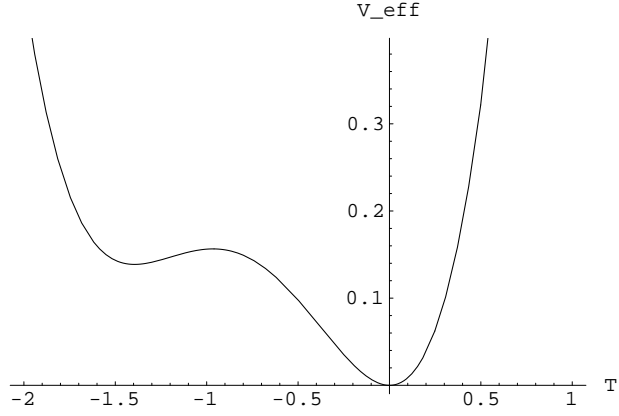}}
	\end{center}
	\caption{The cubic VSFT effective potential without gauge fixing at level (2,4).}
	\label{fig:VSFTpot2}
\end{figure}

Up to level (2,4), the GSO($+$) fields can be integrated out exactly. 
In the Siegel gauge $v_7=v_8=0$, the resulting effective potential for $t$ becomes
\begin{eqnarray}
V^{(\frac{1}{2},1)}_{\mathrm{eff}}(t)&\equiv& -\frac{S^{(\frac{1}{2},1)}_{V,\mathrm{eff}}}{
\kappa_0V_{10}(q_1^2/2)^3}=\frac{t^2(16+9\epsilon t)^2}{256\epsilon^2}, \label{eq:VD} \\
V^{(2,4)}_{\mathrm{eff}}(t)&\equiv& -\frac{S^{(2,4)}_{V,\mathrm{eff}}}{\kappa_0V_{10}(q_1^2/2)^3}= 
\frac{t^2 (96237504 + 119417628 \epsilon t + 37335269 \epsilon^2t^2)}{127993536\epsilon^2}. 
\label{eq:VE} 
\end{eqnarray}
Note that the potential is no longer an even function of $t$ as a consequence of the 
presence of $\Qev$. From the profiles shown in Fig.\ref{fig:VSFTpot1}, 
it is clear that there are two translationally invariant solutions at each level, 
one of which (maximum) would correspond to the unstable D9-brane, 
while the other (minimum) to `another tachyon vacuum' with vanishing 
energy density. If we did not impose any gauge-fixing condition, 
we would obtain the effective potential shown in Fig.~\ref{fig:VSFTpot2} at level (2,4). 
In this potential there is no clear distinction between the maximum and the 
non-trivial minimum. 
Hence we proceed by choosing the Siegel gauge. 

At level (2,6) we can no longer analytically integrate out the massive fields. 
Instead of constructing the effective potential numerically, 
we solve the full set of equations of motion including that for $t$. 
In the Siegel gauge, we have found four real solutions. The field values and 
the potential height for each solution are shown in Table~\ref{tab:sols}. 
\begin{table}[htbp]
	\begin{center}
	\scalebox{0.95}[0.95]{
	\begin{tabular}{|c||r|r|r|r|r|r|}
	\hline
	 & \multicolumn{2}{|c|}{level (2,4)} & \multicolumn{4}{|c|}{level (2,6)} \\
	\cline{2-7}
	 & \multicolumn{1}{|c|}{minimum} & maximum & solution(1) & solution(2) & solution(3) & solution(4) \\
	\hline\hline
	$\epsilon t$ & $-1.58654$ & $-0.812353$ & $-1.17746$ & $-1.27738$ & $-0.602025$ & $-2.21251$ \\
	\hline
	$\epsilon^2 u$ & $-1.38402$ & 0.160340 & 0.293931 & 0.475872 & $-0.554083$ & 2.04485 \\
	\hline
    $\epsilon^2 v_1$ & $-1.02701$ & $-0.486343$ & $-0.483380$ & $-0.309084$ & 0.528839 & $-0.948461$ \\
	\hline
    $\epsilon^2 v_2$ & 0.177618 & 0.0701144 & $0.207129$ & 0.156708 & $-0.116926$ & $-0.660358$ \\
	\hline
    $\epsilon^2 v_3$ & $-0.330186$ & $-0.255363$ & $-0.311352$ & $-0.174039$ & $-0.0944988$ & 0.852255 \\
	\hline
    $\epsilon^2 v_4$ & $-0.00116238$ & $-0.139197$ & $-0.161496$ & $-0.0484503$ & 0.327373 & 0.260061 \\
	\hline
    $\epsilon^2 v_5$ & 1.23853 & 0.249494 & 0.132412 & 0.0669892 & 0.0981466 & $-0.207735$ \\
	\hline
    $v_6$ & 0 & 0 & 0 & 0 & 0 & 0 \\
	\hline
    $v_9$ & 0.276184 & $-0.463130$ & 0.220859 & 0.988841 & $-3.10354$ & $-2.24061$ \\
	\hline\hline
    $\epsilon^4 V_{\mathrm{min}}$ & 0.0148204 & 0.123052 & 0.165213 & 0.175943 & $-0.0233896$ & $-0.625412$ \\
    \hline
	\end{tabular}}
	\end{center}
	\caption{The vacuum expectation values of the fields and the height of the potential 
	for the Siegel gauge solutions.}
	\label{tab:sols}
\end{table}
Comparing them with the level (2,4) solutions, we expect that the solution~(1) 
would correspond to the maximum of the potential. 
However, it seems that there is no candidate solution for the minimum: 
The energy of the solution~(3) is almost zero, but the vev of $t$ is unaccetably too small. 
Therefore, although the seemingly desirable double-well potential was obtained at low levels, 
this success may not continue to level (2,6) or higher.

\subsection{Non-polynomial vacuum superstring field theory}
\begin{figure}[htbp]
	\begin{center}
    \scalebox{0.7}[0.7]{\includegraphics{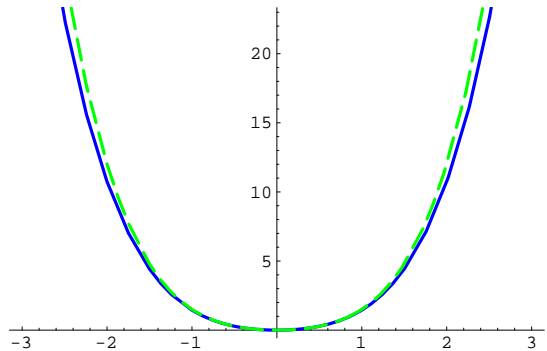}}
	\end{center}
	\caption{The non-polynomial VSFT effective potential in the Siegel gauge  
	at levels (0,0) (dashed line) and ($\frac{3}{2},3$) (solid line).}
	\label{fig:VSFTpot3}
\end{figure}

We also examine the vacuum superstring field theory action 
based on the Berkovits' formulation. It was argued in~\cite{KMS,0208009} that 
the action around the tachyon vacuum should be given by simply replacing the 
kinetic operator $\widehat{Q}_B$ in~(\ref{eq:WZW2}) with $\widehat{\cQ}$, 
\begin{equation}
S_V=\frac{\kappa_0}{2}\sum_{M,N=0}^{\infty}\frac{(-1)^N}{(M+N+2)!}
\left({M+N \atop N}\right)\mathrm{Tr}\bllk\left(\widehat{\cQ}\widehat{\Phi}\right)
\widehat{\Phi}^M\left(\widehat{\eta}_0\widehat{\Phi}\right)\widehat{\Phi}^N
\brrk. \label{eq:B-VSFTaction}
\end{equation}
Let us first consider terms with $M+N\ge 1$. 
Since the conformal transformations of $c(\pm i)$ and $\gamma(\pm i)$ give rise to 
vanishing factors of $(f^{(n)\prime}_k(\pm i))^h$ with $h<0$, 
$\widehat{\cQ}$ reduces to $\widehat{Q}_2=-\oint\frac{dz}{2\pi i}b\gamma^2(z)\otimes\sigma_3$, 
at least for Fock space states $\widehat{\Phi}$. Incidentally, this is reminiscent of  
the pregeometric action proposed in~\cite{pregeo}. For the quadratic vertex ($M=N=0$) 
$f^{(2)\prime}_k(\pm i)$ is finite, so that the midpoint insertions can survive. 
From the above considerations, one sees that the $\zetto_2$-symmetry breaking effect 
(\textit{i.e.} $\Qev$) could come only from the GSO($+$)/GSO($-$) transition vertex 
\begin{equation}
\llk (\Qev\Phi_+)(\eta_0\Phi_-) \rrk. \label{eq:VF}
\end{equation}
However, the actual calculations show that all the above $\zetto_2$-breaking 
terms vanish for level-2 string field (in the Feynman-Siegel gauge)
\begin{eqnarray}
\Phi_+&=&a\ \xi\partial\xi c\partial^2 ce^{-2\phi}+e\ \xi\eta+f\ \xi ce^{-\phi}G^{\mathrm{m}},\nonumber \\
\Phi_-&=&t\ \xi ce^{-\phi}+k\ \xi c\partial^2(e^{-\phi})+l\ \xi c\partial^2\phi e^{-\phi} \label{eq:VG} \\
& &{}+m\ \xi cT^{\mathrm{m}}e^{-\phi}+n\ \xi\partial^2ce^{-\phi}+p\ \xi\partial\xi\eta ce^{-\phi}. \nonumber
\end{eqnarray} 
As a result, the effective potential for the lowest mode $t$ becomes
left-right symmetric as shown in Fig.~\ref{fig:VSFTpot3}. 
To make matters worse, there exist no real solutions other than $\widehat{\Phi}=0$. 
From a lot of examples we have learned that in the successful level truncation calculations 
the correct (expected) qualitative behavior is reproduced already at the lowest level, 
and the contributions from higher-level states only give small corrections to it. 
If we assume this empirical law in this case as well, 
we should attribute the above unwelcome result to the fact that there is no 
$t^3$ term in the action at level (0,0). 
However, to make the 3-string vertex 
$ \llk (\Qev\Phi_-)\Phi_-(\eta_0\Phi_-) \rrk $
non-vanishing for $\Phi_-=t \xi ce^{-\phi}$ we must modify the precise form of $\Qev$. 
In particular, the insertion of the negative-dimensional operators to the 
open string midpoint does not fulfill this purpose. However, 
no alternatives are available now since it is very difficult to construct nilpotent $\widehat{\cQ}$ 
with $\Qev\neq 0$.

\sectiono{Summary and Discussion}\label{sec:summary}
In the first half (sections 2--4) of this paper, we have examined the  
component structure of the superstring field theory action in some detail. 
We have explicitly shown that in modified cubic theory 
the correct Maxwell action~(\ref{eq:Maxwell}) 
and tachyon action~(\ref{eq:YV}) are obtained after integrating out 
the auxiliary fields. We have specified the precise way of fixing the sign ambiguities 
arising in the conformal factors of the GSO($-$) components in the case of 
the UHP representation of the string vertices. 
Furthermore, we have discussed the conditions on the string fields which guarantee the 
reality of the action, for all of modified cubic, Witten's cubic 
and Berkovits' non-polynomial superstring field theories, using the CFT method. 

The latter half is devoted to the level truncation analysis of the tachyon 
condensation problems. In modified cubic superstring field theory, 
though the tachyon potential on a non-BPS D-brane as well as its minimum is 
well constructed up to level (2,6) 
in the Feynman-Siegel gauge, its extension to higher levels may 
be subtle. We have not found any gauge choice which seems particularly better than the 
Feynman-Siegel gauge. 
Restricting ourselves to the lowest level ($\frac{1}{2},1$), 
we have obtained the static kink solution representing a BPS D-brane of one lower dimension, 
whose energy density is remarkably close to the expected D-brane tension 
($\sim$101\%). We have also verified that no such non-perturbative vacuum as was found 
in~\cite{AMZssb} in the theory with the picture-changing operator $\cZ$ 
exists on a BPS D-brane if we employ $Y(i)Y(-i)$ as the picture-changing operator 
which is preferable to $\cZ$. Lastly, we have investigated 
whether vacuum superstring field theory with 
the pure-ghost kinetic operator~(\ref{eq:VSFTop}) can support the correct (expected) solutions 
in the level truncation scheme. Unfortunately, we have obtained disappointing 
results in both of cubic and Berkovits' non-polynomial theories.  
\bigskip

In the study of tachyon condensation in modified cubic superstring field theory, 
we have found the following unusual features: 
(i) The potential depth and the kink tension are very close to 
the expected values already at the lowest level $\left( \frac{1}{2},1\right)$, 
(ii) The vacuum energy does not seem to improve regularly as the truncation level is increased, 
(iii) The tachyon vacuum is not reached in the Feynman-Siegel gauge 
at level $\left( \frac{5}{2},5\right)$~\cite{Raeymaekers}.
These are in contrast with the results obtained in bosonic and Berkovits' 
theories~(see \cite{GR,DeSmet}). We consider these behaviors should be attributed to the 
unconventional choice (0-picture) of field variables. More precisely, 
we would like to suggest the following interpretation: 
Let us recall that a solution $\Phi_0$ in Berkovits' superstring field theory and 
a solution $A_0$ in modified cubic theory which share the same physical content 
are formally related through the map $A_0=e^{-\Phi_0}Q_Be^{\Phi_0}$ (see~\cite{0208009} for details). 
Then, the low-lying fields in $A_0$ would receive contributions from various higher modes in $\Phi_0$, 
because the $*$-product mixes fields of different levels. 
Furthermore, since $b_0$ is \textit{not} a derivation of the $*$-algebra, a Siegel gauge solution 
$\Phi_0$ in Berkovits' theory does not in general map to a Siegel gauge solution $A_0$ in 
modified cubic theory. Given that the Siegel gauge solution for the tachyon vacuum 
shows the `regular' behavior in Berkovits' theory~\cite{DeSmet}, the above consideration 
may give a possible explanation 
for all the strange behaviors (i)--(iii) of modified cubic theory, 
though we cannot prove it at all. 
\medskip

In light of the results obtained in the level truncation analysis of 
vacuum superstring field theory, 
it seems that the pure-ghost kinetic 
operator~(\ref{eq:VSFTop}) fails to describe the theory 
around the tachyon vacuum. It is even possible that the pure-ghost ansatz for 
the kinetic operator is too simple to correctly reproduce the 
complicated D-brane spectrum of type II superstring theory. 
If this is indeed true, we have to look for a matter-ghost mixed kinetic operator 
which is suitable for the description of the tachyon vacuum. 
This would require us to make a fresh start 
for the construction of vacuum superstring field theory.


\section*{Acknowledgements}
I would like to thank T. Eguchi, T. Kawano, I. Kishimoto and K. Sakai for useful discussions,  
and Y. Matsuo and W. Taylor for encouragements. 
I also thank J. Raeymaekers for e-mail correspondence. 
This work is supported by JSPS Research Fellowships for Young Scientists. 
\nopagebreak

\section*{Appendices}
\renewcommand{\thesection}{\Alph{section}}
\setcounter{section}{0}

\sectiono{Cubic Action at Level (2,6)}\label{sec:AppA}
The cubic action truncated at level (2,6) is found to be (in units where $\ap =1$)
\begin{eqnarray}
f&=&-\frac{S}{\tilde{\tau}_9V_{10}}=-\frac{2\pi^2g_o^2S}{V_{10}}
\equiv f_{\mathrm{quad}}+f_{\mathrm{cubic}}; \label{eq:ApA} \\
f_{\mathrm{quad}}&=&-2\pi^2 \biggl[\frac{1}{4}t^2+\frac{1}{2}u^2 +\sqrt{2} u v_1 
+v_1^2+\frac{15}{8}v_2^2-\frac{1}{\sqrt{2}}u v_3+2 v_1 v_3 \nonumber \\
& &{}+\frac{1}{4}v_3^2-2\sqrt{2}u v_4
-8 v_1 v_4-4 v_3 v_4+\frac{77}{8}v_4^2+2\sqrt{2}u v_5+6 v_1 v_5+v_3v_5 \nonumber \\
& &{}-13 v_4v_5+\frac{11}{2}v_5^2
-\frac{15}{2}v_2v_6+\frac{15}{2}v_4v_6-5 v_5v_6+\frac{5}{2}v_6^2
+\frac{1}{\sqrt{2}}uv_7+v_1v_7 \nonumber \\
& &{}-\frac{1}{2}v_3v_7-2 v_4v_7+2v_5v_7+3v_3v_8-5v_4v_8+2v_5v_8+v_7v_8
+\frac{1}{\sqrt{2}}u v_9 \nonumber \\
& &{}+2v_1v_9-\frac{15}{4}v_2v_9
+v_3v_9-\frac{5}{4}v_4v_9+v_5v_9
+v_7v_9+v_8v_9\biggr], \label{eq:ApB1} \\
f_{\mathrm{cubic}}&=&-2\pi^2\bigg[ \frac{9\sqrt{2}}{16}t^2 u+\frac{9}{8}t^2v_1
-\frac{25}{32}t^2 v_2-\frac{9}{16}t^2 v_3-\frac{59}{32}t^2 v_4
+\frac{43}{24}t^2 v_5+\frac{40}{9}\sqrt{\frac{2}{3}}u v_6^2 \nonumber \\
& &{}+\frac{80}{9\sqrt{3}}v_1v_6^2-\frac{20}{9\sqrt{3}}v_2v_6^2-\frac{40}{9\sqrt{3}}v_3v_6^2-\frac{1180}{81\sqrt{3}}
v_4v_6^2+\frac{3440}{243\sqrt{3}}v_5v_6^2+\frac{2}{3}t^2 v_7 \nonumber \\
& &{}+\frac{1280}{243\sqrt{3}}v_6^2v_7+\sqrt{3}u^2v_9+\frac{70}{9}\sqrt{\frac{2}{3}}u v_1v_9+\frac{86}{9\sqrt{3}}v_1^2v_9
-\frac{25}{3\sqrt{6}}uv_2v_9-\frac{875}{81\sqrt{3}}v_1v_2v_9 \nonumber \\
& &{}+\frac{4435}{648\sqrt{3}}v_2^2v_9+\frac{5}{9}\sqrt{\frac{2}{3}}uv_3v_9+\frac{350}{243\sqrt{3}}
v_1v_3v_9-\frac{125}{162\sqrt{3}}v_2v_3v_9-\frac{37}{18\sqrt{3}}v_3^2v_9 \nonumber \\
& &{}-\frac{193}{9\sqrt{6}}uv_4v_9
-\frac{6755}{243\sqrt{3}}v_1v_4v_9+\frac{4825}{324\sqrt{3}}v_2v_4v_9
-\frac{965}{486\sqrt{3}}v_3v_4v_9+\frac{39809}{1944\sqrt{3}}v_4^2v_9 \nonumber \\
& &{}+\frac{86}{9}\sqrt{\frac{2}{3}}uv_5v_9+\frac{6020}{243\sqrt{3}}v_1v_5v_9
-\frac{1075}{81\sqrt{3}}v_2v_5v_9+\frac{430}{243\sqrt{3}}v_3v_5v_9
-\frac{979}{27\sqrt{3}}v_4v_5v_9 \nonumber \\
& &{}+\frac{4082}{243\sqrt{3}}v_5^2v_9
+\frac{8}{3}\sqrt{\frac{2}{3}}uv_7v_9+\frac{1552}{243\sqrt{3}}v_1v_7v_9-\frac{100}{27\sqrt{3}}
v_2v_7v_9+\frac{40}{81\sqrt{3}}v_3v_7v_9 \nonumber \\
& &{}-\frac{772}{81\sqrt{3}}v_4v_7v_9+\frac{688}{81\sqrt{3}}v_5v_7v_9+\frac{16}{9}
\sqrt{\frac{2}{3}}uv_8v_9+\frac{32}{9\sqrt{3}}v_1v_8v_9-\frac{200}{81\sqrt{3}}v_2v_8v_9 \nonumber \\
& &{}+\frac{80}{243\sqrt{3}}v_3v_8v_9-\frac{1544}{243\sqrt{3}}
v_4v_8v_9+\frac{1120}{243\sqrt{3}}v_5v_8v_9+\frac{256}{81\sqrt{3}}v_7v_8v_9\biggr]. \label{eq:ApB2}
\end{eqnarray}
As a verification of our result, 
let us compare it with the results of refs.~\cite{Raeymaekers,ABKM1}. 
Our function $f$~(\ref{eq:ApA}) up to level (2,4) precisely agrees with that of 
Raeymaekers~\cite{Raeymaekers}, if we make the replacements 
\begin{eqnarray}
& & v_7\to 0, \quad v_8\to 0, \quad (\mbox{Feynman-Siegel gauge}) \label{eq:ApC} \\
& & v_6\to -v_6, \quad u\to\frac{u}{\sqrt{2}}, \quad t\to\frac{t}{2}, \nonumber
\end{eqnarray}
and then $v_9\to v_7$. 
Ours, however, does \textit{not} coincide with the result of Aref'eva \textit{et al.} 
(version 3 of~\cite{ABKM1}) even after setting 
\begin{equation}
v_6\to -v_6, \quad v_9\to 0, \quad u\to\frac{u}{\sqrt{2}}, \quad t\to\frac{t}{2}. \label{eq:ApD}
\end{equation}
In view of our and Raeymaeker's results, 
the $-$ sign in front of the parenthesis in the last line of eq.(3.3) 
of (version 3 of) \cite{ABKM1} should be $+$. If so, ours and theirs agree with each other. 
\medskip

The cubic vacuum superstring field theory action truncated up to level (2,6) is, 
after the rescaling $\cA_{\pm}\to (q_1^2/2)\cA_{\pm}$ mentioned in subsection~\ref{subsec:cubVSFT}, 
given by 
\begin{eqnarray}
S_V&=&\kappa_0\left(\frac{q_1^2}{2}\right)^3V_{10}\left(-\tilde{f}_{\mathrm{quad}}-\frac{1}{2\pi^2}
f_{\mathrm{cubic}}\right), \label{eq:ApE} \\
-\tilde{f}_{\mathrm{quad}}&=&\frac{1}{2}u^2+\sqrt{2}uv_1+v_1^2+\frac{15}{8}v_2^2
-\frac{1}{\sqrt{2}}uv_3+2v_1v_3+\frac{1}{4}v_3^2-2 \sqrt{2}uv_4-8v_1v_4 \nonumber \\
& &{}-4 v_3v_4+\frac{77}{8}v_4^2+2\sqrt{2}uv_5+6v_1v_5+v_3v_5-13v_4v_5
+\frac{11}{2}v_5^2+\frac{1}{\sqrt{2}}uv_7 \nonumber \\
& &{}+v_1v_7-\frac{1}{2}v_3v_7-2v_4v_7+2 v_5v_7
+3v_3v_8-5v_4v_8+2v_5v_8+v_7v_8 \nonumber \\ & &{}+\frac{1}{\epsilon}
\left(\sqrt{2}tu+2tv_1-tv_3-\frac{5}{2}tv_4+3tv_5+tv_7\right), \nonumber 
\end{eqnarray}
where $\epsilon =q_1\varepsilon_r$ and $f_{\mathrm{cubic}}$ is the same as in~(\ref{eq:ApB2}). 

\sectiono{Technical Remarks about the Correlators and the Conformal Transform 
of $e^{ikX}$}\label{sec:AppB}
The fact that each $X^{\mu}$ contains both left- and right-movers makes 
the computations of the correlators including $X^{\mu}$ in the open string case complicated. 
In the presence of the open string boundary, 
the OPE between two $X$'s inserted in the 
interior of the world-sheet becomes~\cite{Polchinski}
\begin{equation}
X^{\mu}(z,\bar{z})X^{\nu}(w,\bar{w})\sim -\frac{\ap}{2}\eta^{\mu\nu}\ln |z-w|^2-\frac{\ap}{2}
\eta^{\mu\nu}\ln |z-\bar{w}|^2. \label{eq:ApM}
\end{equation}
Hence, when two $X$'s are inserted on the boundary ($z=\bar{z},w=\bar{w}$) we should have 
\begin{equation}
X^{\mu}(z)X^{\nu}(w)\sim -2\ap\eta^{\mu\nu}\ln (z-w), \label{eq:ApN}
\end{equation}
where $z,w$ are real numbers satisfying $z>w$. The $XX$ OPE appearing 
in~(\ref{eq:ZT}) should be understood this way. 

If we want to calculate the string field theory vertices for string fields 
having non-zero momenta, we must compute the conformal transformations 
$f^{(n)}_k\circ e^{ikX}$ and the correlators 
$\langle e^{ik_1X(z_1)}\cdots e^{ik_nX(z_n)}\rangle_{\mathrm{UHP}}$. 
Since the world-sheet scalars $X^{\mu}$ are 
\textit{bosonic} variables, two exponentials of $X$ must \textit{commute} with 
each other without any phase factor irrespective of the values of momenta. 
For the OPE to be consistent with this commutation rule, we must have 
\begin{eqnarray}
& &:e^{ik_1X(z_1)}::e^{ik_2X(z_2)}:\ \sim |z_1-z_2|^{2\ap k_1\cdot k_2}
:e^{ik_1X(z_1)+ik_2X(z_2)}:, \label{eq:ApO} \\
& &\partial X^{\mu}(z):e^{ikX(w)}:\ =\ :e^{ikX(w)}:\partial X^{\mu}(z)
\sim -\frac{2\ap ik^{\mu}}{z-w}:e^{ikX(w)}:, \nonumber 
\end{eqnarray}
on the boundary. In general, the $n$-point correlator among $e^{ik_jX(z_j)}$ becomes 
\begin{equation}
\langle e^{ik_1X(z_1)}\cdots e^{ik_nX(z_n)}\rangle_{\mathrm{UHP}}=\left(\prod_{i>j}
|z_i-z_j|^{2\ap k_i\cdot k_j}\right)(2\pi)^{10}\delta^{10}\left(\sum k_i\right). \label{eq:ApP}
\end{equation}
From the remark in the last paragraph, 
it would be natural to consider that the conformal factor of $e^{ikX}$ 
contains both contributions from 
holomorphic and antiholomorphic sides. Then, since $e^{ikX}$ is a primary field of 
conformal weight $\ap k^2$, 
the conformal transform of $e^{ikX}$ by $f$ should be given by 
\begin{equation}
f\circ e^{ikX(z)}=|f^{\prime}(z)|^{\ap k^2}e^{ikX(f(z))}. \label{eq:ApR}
\end{equation}
Otherwise, the phase of the conformal factor would be ill-defined for a general value of momentum. 
In the particular case of $f=I$ (inversion), we have 
\begin{equation}
I\circ e^{ikX(z)}=|z^{-2}|^{\ap k^2}e^{ikX(-1/z)}. \label{eq:ApT}
\end{equation}
Finally, the hermitian conjugation of $e^{ikX(z)}$ is defined to be 
\begin{equation}
\left( e^{ikX(z)}\right)^{\dagger}=|z^{-2}|^{\ap k^2}e^{-ikX(1/z^*)}, \label{eq:ApS}
\end{equation}
in accordance with (\ref{eq:ApT}). Note that there is no difference between $z^{2}$ and 
$|z^2|$ for real $z$. 
We have performed all the calculations in the text according to the above rules. 


\end{document}